\documentclass[10pt]{article}
\usepackage{epsf,amsfonts,hyperref}
\usepackage{graphicx}
\renewcommand{\appendix}[1]{
    \addtocounter{section}{1}
    \setcounter{equation}{0}
    \renewcommand{\thesection}{\Alph{section}}
    \section*{Appendix \thesection\protect\indent #1}
    \addcontentsline{toc}{section}{Appendix \thesection\ \ \ #1}
}
\newcommand\encadremath[1]{\vbox{\hrule\hbox{\vrule\kern8pt
\vbox{\kern8pt \hbox{$\displaystyle #1$}\kern8pt}
\kern8pt\vrule}\hrule}}
\def\enca#1{\vbox{\hrule\hbox{
\vrule\kern8pt\vbox{\kern8pt \hbox{$\displaystyle #1$}
\kern8pt} \kern8pt\vrule}\hrule}}

\newcommand\figureframex[3]{
\begin{figure}[bth]
\hrule\hbox{\vrule\kern8pt
\vbox{\kern8pt \vbox{
\begin{center}
{\mbox{\epsfxsize=#1.truecm\epsfbox{#2}}}
\end{center}
\caption{#3}
}\kern8pt}
\kern8pt\vrule}\hrule
\end{figure}
}
\newcommand\figureframey[3]{
\begin{figure}[bth]
\hrule\hbox{\vrule\kern8pt
\vbox{\kern8pt \vbox{
\begin{center}
{\mbox{\epsfysize=#1.truecm\epsfbox{#2}}}
\end{center}
\caption{#3}
}\kern8pt}
\kern8pt\vrule}\hrule
\end{figure}
}

\newcommand{\eq}[1]{Eq.~(\ref{#1})}

\newcommand{\beq}{\begin{equation}}
\newcommand{\eeq}{\end{equation}}
\newcommand{\bea}{\begin{eqnarray}}
\newcommand{\eea}{\end{eqnarray}}

\renewcommand{\and}{{\qquad {\rm and} \qquad}}

\newcommand{\virg}{{\qquad , \qquad}}

 \newcommand{\Tr}{{\,\rm Tr}\:}

\newcommand{\Res}{\mathop{\,\rm Res\,}}

\newcommand{\Pint}{{\int\kern -1.em -\kern-.25em}}

\newcommand{\Pol}{\mathop{\mathrm{Pol}}}
\newcommand{\CZ}{{\cal Z}}
\newcommand{\CN}{{\cal N}}

\newcommand{\moy}[1]{\left<{#1}\right>}

\renewcommand{\thesection}{\arabic{section}}

\makeatletter
\@addtoreset{equation}{section}
\makeatother
\newtheorem{theorem}{Theorem}[section]

\newtheorem{remark}{Remark}[section]
\newtheorem{proposition}{Proposition}[section]
\newtheorem{lemma}{Lemma}[section]
\newtheorem{corollary}{Corollary}[section]
\newtheorem{definition}{Definition}[section]
\def\br{\begin{remark}\rm\small}
\def\er{\end{remark}}
\def\bt{\begin{theorem}}
\def\et{\end{theorem}}
\def\bd{\begin{definition}}
\def\ed{\end{definition}}
\def\bp{\begin{proposition}}
\def\ep{\end{proposition}}
\def\bl{\begin{lemma}}
\def\el{\end{lemma}}
\def\bc{\begin{corollary}}
\def\ec{\end{corollary}}
\def\beaq{\begin{eqnarray}}
\def\eeaq{\end{eqnarray}}
\newcommand{\proof}[1]{{\noindent \bf proof:}\par
{#1} $\square$}

\begin{document}
%=============================Page de titre===============

\sloppy

%\maketitle

\pagestyle{empty}
\hfill 
\addtolength{\baselineskip}{0.20\baselineskip}
\begin{center}
\vspace{26pt}
\begin{center}
{\large \bf {Non-homogenous disks in the chain of matrices}}
\newline
\end{center}
\vspace{26pt}

{\sl N.\ Orantin$^a$, A.\ Veliz-Osorio$^b$}\hspace*{0.05cm}\footnote{ E-mail: norantin@math.ist.utl.pt, alvaro.osorio@ist.utl.pt}\\
\vspace{6pt}
$ ^a$ CAMSD, Departamento de Matem\'{a}tica, Instituto Superior T\'{e}cnico\\
Av. Rovisco Pais 1, 1049-001 Lisboa, Portugal\\ 
\bigskip
$ ^b$ Departamento de F\'{i}sica, Instituto Superior T\'{e}cnico\\
Av. Rovisco Pais 1, 1049-001 Lisboa, Portugal

\end{center}

\vspace{20pt}
\begin{center}
{\bf Abstract}: We investigate the generating functions of multi-colored discrete disks with non-homogenous boundary condition in the context of the Hermitian multi-matrix model where the matrices are coupled in an open chain. We show that the study of the spectral curve of the matrix model allows to solve a set loop equations to get a recursive formula computing mixed trace correlation functions to leading order in the large matrix limit.
\end{center}

%\newpage
\vspace{26pt}
\pagestyle{plain}
\setcounter{page}{1}
%*********************************************************************
%==================== ARTICLE ========================================
%*********************************************************************

\section{Introduction}

The theory of random matrices has many ramifications in different fields of mathematics and physics. In the recent years, following the work of Eynard \cite{E1MM}, matrix models have
seen significant advances especially through their applications to enumerative geometry. The enumeration of surfaces can indeed be addressed by the evaluation of the result of a saddle point approximation on integrals defined over a set of random matrices as the size of the matrices gets very large. To be more precise, the result of such an approximation can be seen as the generating function of a set of fat graphs composed of ribbons glued together along vertices \cite{BIPZ}. This set of fat graphs is in bijection with the set of discrete surfaces, or maps as denoted by combinatorists, i.e. surfaces composed of polygons glued by their edges. The matrix integrals can thus be seen, in this  large matrix limit, as the generating function of a set of maps with a weight depending on the probability measure put on the set of matrices under study \cite{ambjornrmt,David,Kazakov}. Such objects are of prime interest not only to mathematicians but also to physicists for investigating the behavior of statistical systems on random surfaces (or lattices) as well as for understanding quantization of two dimensional gravity or topological string theories from a discrete point of view. Discrete surfaces also provide a very good toy model for developing some intuition to be applied later to string and gauge theories.

From this perspective,  a lot of progress have been made by Eynard et al. leading to powerful universal technics for the computation of generating functions of maps of different types. Among the results obtained, one finds:

\begin{itemize}

\item The enumeration of maps of arbitrary topology, i.e. arbitrary genus and number of boundaries, has been achieved in \cite{ACKM,ec1loopF,E1MM};

\item The enumeration of bicolored maps of arbitrary genus and arbitrary boundary conditions has been completed in the series of papers \cite{eynm2m,EO1,CEO,EKK,EOsymmetry,EOallmixed};

\item The enumeration of bicolored maps drawn on non-orientable surfaces has also been solved in \cite{ChEynbeta,EM,CEM1,CEM2,BEMPF} with homogenous boundary conditions;

\item The enumeration of maps whose faces are colored with more than two colors has been partly solved in \cite{Echain,EPF}. In particular, the generating functions of maps with homogenous boundary conditions are known whatever their genus is.

\end{itemize}

One big outcome of these works is the universality of the solution found for all these problems. Indeed, for each of the aforementioned enumerative problems, the generating function of maps with homogenous boundary conditions are given by a unique universal inductive formula on the Euler characteristic of the surfaces enumerated. Even better is the fact that this inductive formula, sometimes called topological recursion, goes beyond the field of random matrices and enumeration of maps. The topological recursion \cite{EOinvariants,EOrev} seems to be a universal solution to different problems of enumerative geometry consisting in the computation of the volume of the moduli space of Riemann surfaces with respect to different measures opening the way for new insights in mathematics and topological string theories even when this measure does not localize on discrete surfaces.

The topological recursion was inspired by the computation of generating functions of open surfaces with homogenous boundary conditions in the different matrix models encountered. The next step in fully understanding the enumeration of maps in general is to understand the action of boundary operator, that is compute the generating function of surfaces with boundaries along which the boundary condition changes. This has been understood in the hermitian two matrix model case where it has been possible to compute the generating function of surfaces composed of polygons of two different colors with arbitrary boundary conditions \cite{EOallmixed}. The result points towards a generalization of the topological recursion for these new observables. Going further requires the study of the so-called chain of hermitian matrix model which generates maps colored with an arbitrary number of colors. Even if the homogenous boundary condition case is understood \cite{EPF}, there are only a few result concerning changes of boundary conditions \cite{Echain}.

In these notes, we present a recursive method, similar to the topological recursion, which allows to compute the generating function of discs with mixed boundary conditions. 
In theorem \ref{threc} and corollary \ref{correc} we prove that the corresponding correlation function $\left<\Tr \left[ {1 \over x_1-M_1} {1 \over x_2-M_2} \dots {1 \over x_\CN-M_\CN} \right]\right>$ can be computed to leading order in the large matrix limit in terms of the spectral curve of the model. The formula takes a very nice form which seems to follow from the all possible degeneracies of the surfaces generated.
From the two matrix model experience, it is probably the building block for all the other generating functions and probably one of the most important objects of the theory. We expect to be able to use the techniques developed in this paper to compute any observable of this model.

The computation of these disk amplitudes has thus many interesting features. 
First of all, it gives an efficient method for computing the generating functions of discs composed of colored polygons with mixed-boundary conditions. In addition, we are convinced that the result as well as the methods used in its derivation are  generalizable to the computation of the generating function of surfaces of arbitrary topology. From this perspective, these notes  should provide the building block for the computation of all observable of the chain of matrices model. This result is also important for better understanding conformal field theories. Indeed, when going to a particular limit in the parameter space of matrix models, one reaches critical regimes described by minimal models, a particular type of rational conformal field theories. 
The knowledge of boundary operators in the matrix model setup thus also gives access to boundary operator in such rational conformal field theories.
Finally, the form of the solution points towards a generalization outside of the matrix model's setup generalizing the topological recursion formalism leading to the knowledge of new open amplitudes in topological string theories or may be new relative Gromov-Witten invariants.

\bigskip

This paper is organized as follows:
\begin{itemize}

\item In section 2, we introduce the model as well as the notations required in the following and present the result of these notes;

\item In section 3, we remind how algebraic geometry appears in this context through the so-called loop equations and introduce all the notations
and basic knowledge further needed.

\item In section 4, we derive and solve a closed set of loop equation satisfied by the leading order mixed correlations functions.

\item In section 5, we specialize this result to the chain of 3 matrices and give explicit examples.

\item Section 6 is devoted to the conclusion and a short discussion on the perspectives.

\end{itemize}

\section{The open chain of matrices}

\subsection{The model}

We want to generate discrete surfaces, i.e. surfaces composed of polygons glued by their edges. In addition, we want to include some additional information on these random surfaces such as a spin structure or some matter. For this purpose, each polygon carries one color labeled by an index ranging from 1 to $\CN$.

Generating functions for such discrete colored surfaces can be obtained as correlation functions of a multi-matrix model.
The latter studies the distribution of matrix elements  with respect to a probability measure $d\mu$ on the set $\left(H_N\right)^{\cal N}$ of ${\cal N}$-uple of hermitian matrices of respective size $N \times N$ defined as
\beq
d\mu\left(M_1,\dots,M_{\cal N}\right)  := dM_1 dM_2 \dots dM_{\CN} \exp\left(- {N \over T} \Tr\left[\sum_{k=1}^{\CN} V_k(M_k) - \sum_{k=2}^{\CN} c_{k-1,k} M_{k-1} M_k\right] \right)
\eeq
where $M_i$ is a hermitian matrix of size $N$, $dM_i$
is the product of Lebesgue measures of the real components of $M_i$ and
\beq
V_k(x) =  \sum_{i=2}^{d_k+1} {g_i^{(k)} \over i} x^{i}
\eeq
are polynomial potentials of respective degrees $d_k+1$.
The partition function of this model is defined as
\beq\label{defpartition}
\CZ := \int_{\left(H_N\right)^{\cal N}} d\mu\left(M_1,\dots,M_{\cal N}\right).
\eeq
In this paper, we interpret this random matrix integral as a generating function of discrete surfaces. This interpretation is exact if one considers a formal version of this integral, which is defined as the result of a saddle point approximation of this one\footnote{This formal integral is
defined as a formal series in $T$  and does not need to converge. More precisions about this definition can be found in \cite{Emap}}. In the following, all the matrix integrals encountered are formal in this sense.

In a nutshell, the combinatorial
interpretation of such formal matrix integral is obtained by expanding the non quadratic part of the action
\beq
{\cal S} := \sum_{k=1}^{\CN} V_k(M_k) - \sum_{k=2}^{\CN} M_{k-1} M_k
\eeq
around the extremum $\left\{M_i = 0\right\}_{i=1}^{\cal N}$ and represent the result by Feynman graphs whose edges are ribbons which cannot be twisted.
In the expansion of the action around this saddle, each factor $\Tr M_i^k$ is represented by a $k$-valent vertex of color $i$ whose legs are ribbons.
This peculiarity of the edges puts an ordering on the legs around every vertex. One can thus replace every vertex by a face (or polygon) whose number of edges equals the valence of the initial vertex. Then, each graph is dual to a map, or discrete surface.

Hence, the partition function of the matrix model is defined as the generating function of closed surfaces made of polygons colored
with $\CN$ different colors and glued together following the prescriptions:
\begin{itemize}

\item the polygons of color $k$ have at most $d_k+1$ edges ;

\item two polygons of respective colors $i$ and $j$ can be glued by their edges only.

\end{itemize}
In this representation, the partition function reads
\beq
\CZ = \sum_{v=0}^\infty T^v \sum_{g=0}^\infty N^{2-2g}  \sum_{S \in \mathbb{M}_g(v)} {1 \over \# \hbox{Aut}(S)} \prod_{i,k} \left(g_i^{(k)}\right)^{n_{i,k}(S)} \prod_{i,j} \left(\left[C^{-1}\right]_{i,j}\right)^{\widetilde{n}_{i,j}(S)}
\eeq
where $\mathbb{M}_g(v)$ is the set of orientable genus $g$ maps with $v$ vertices, $\hbox{Aut}(S)$ is the number of automorphism of $S$, $n_{i,k}(S)$ the number of faces of color $k$ with $i$ edges in $S$, $\widetilde{n}_{i,j}(S)$ the number of edges between a face of color $i$ and a face of color $j$ in $S$ and $\left[C^{-1}\right]$ the inverse of the matrix of the bilinear terms of the action
\beq
C :=\left(
\begin{array}{cccccc}
g_2^{(1)}&-c_{1,2}&&&&0\cr
-c_{1,2}&g_2^{(2)}&-c_{2,3}&&&\cr
&-c_{2,3}&g_2^{(3)}&-c_{3,4}&&\cr
&&\ddots&\ddots&\ddots&\cr
&&&-c_{\CN-2,\CN-1}&g_2^{(\CN-1)}&-c_{\CN-1,\CN}\cr
0&&&&-c_{\CN-1,\CN}&g_2^{(\CN)}\cr
\end{array}
\right) .
\eeq

The free energy
\beq
{\cal F}:= \ln \CZ
\eeq
is the generating function of the connected surfaces contributing to $\CZ$.

For any function $f: \left(H_N\right)^{\cal N} \to \mathbb{C}$, one defines a corresponding correlation function by
\beq
\moy{f\left(M_1, \dots,M_{\cal N}\right)}:= {1 \over \CZ} \int_{\left(H_N\right)^{\cal N}}  f\left(M_1, \dots,M_{\cal N}\right) d\mu\left(M_1,\dots,M_{\cal N}\right).
\eeq

The correlation functions of this model can also be interpreted as the generating functions of open surfaces built with the same prescription and the same
measure as the closed ones, i.e. surfaces with boundaries and prescribed boundary conditions on it. Let us briefly recall how to
define generating functions of such surfaces. Consider a correlation function of the form:
\beq
 \left< \Tr \left[\left(M_{i_1}\right)^{j_1} \left(M_{i_2}\right)^{j_2} \dots \left(M_{i_k}\right)^{j_k} \right]\right> := {1 \over \CZ} \int_{\left(H_N\right)^{\cal N}}  \Tr  \left[\left(M_{i_1}\right)^{j_1} \left(M_{i_2}\right)^{j_2} \dots \left(M_{i_k}\right)^{j_k} \right]
d\mu\left(M_1,\dots,M_{\cal N}\right)
\eeq
where $(i_1, i_2, \dots, i_k) \in [1,\CN]^k$ and $(j_1, \dots, j_k)\in \left(\mathbb{N}^*\right)^k$ are arbitrary positive integers.
 The
factor $\Tr \left[\left(M_{i_1}\right)^{j_1} \left(M_{i_2}\right)^{j_2} \dots \left(M_{i_k}\right)^{j_k} \right]$ in the integrand constrains all the generated graphs to
contain one vertex with ${\displaystyle \sum_{l = 1}^k i_l}$ colored legs organized so that the sequence of colors
coincide with the sequence of matrices in $\left(M_{i_1}\right)^{j_1} \left(M_{i_2}\right)^{j_2} \dots \left(M_{i_k}\right)^{j_k}$.
Thus, this correlation function is the generating function of all possible fat graphs that can be glued to this vertex,
i.e. all fat graphs with one boundary constrained by the condition that it must be glued to this vertex.
In the dual representation where k-valent vertices  are replaced by k-gones of the same color, this function
generates all surfaces with one boundary constrained by the same type of gluing condition.

\begin{figure}
  \begin{center}
  \includegraphics[width=5cm,angle=00]{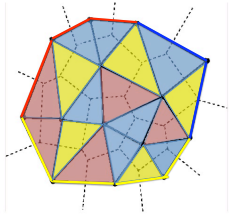}\\
  \end{center}
  \caption{ Example of map contributing to $\left<\Tr M_1^4 M_2^3 M_3^2\right>$ where the type 1 condition (resp. type 2 and type 3) is depicted in yellow (resp. red and blue). It is important recalling that the boundary condition corresponds to the color of the outer face which has been removed and not the color of the inner polygons touching the boundary. The dual fat graph is depicted in dotted lines. }\label{discretefig}
\end{figure}

In order to deal with all the possible exponents $j_l$ at once one introduces the resolvents
\beq\begin{array}{rcl}
W_{i_1,i_2, \dots, i_k}(x_1,x_2, \dots, x_k)&:=&
{1 \over N} {\displaystyle \sum_{j_1, \dots, j_k=1}^\infty} {\left< \Tr \left[\left(M_{i_1}\right)^{j_1} \left(M_{i_2}\right)^{j_2} \dots \left(M_{i_k}\right)^{j_k} \right] \right>
\over x_1^{j_1+1} x_2^{j_2+1} \dots x_k^{j_k+1} }\cr
&:=& {1\over N} \left< \Tr {1 \over x_1 - M_{i_1}} {1 \over x_2 - M_{i_2}} \dots {1 \over x_k - M_{i_k}} \right>\cr
\end{array}
\eeq
as  formal power series in its parameters $x_1,x_2, \dots, x_k$. It is very important to note that these
resolvents are asymptotic series {\em defined only when there arguments are large!}. Indeed,
we will show that these correlation functions are multivalued functions of their variables $x$ in the complex plane\footnote{This property can be seen as the result of the finiteness of the radius of convergency of this series.}  but can be
promoted to monovalued functions on an Riemann surface. However, one must always keep in mind that the
physical quantities are the coefficients of the expansion of these correlation functions around a very particular point.

Using the observation of 't Hooft for QCD \cite{thooft}, one can also select the genus of the generated
surfaces. In order to do so, one can show that these correlation functions are series in ${1\over N^2}$ and that the
order of ${1\over N^2}$ is genus of the generated surface, i.e. one can write:
\beq
W_{i_1,i_2, \dots, i_k}(z_1,z_2, \dots, z_k)
= \sum_{g=0}^{\infty} {1 \over N^{2g}} W_{i_1,i_2, \dots, i_k}^{(g)}(z_1,z_2, \dots, z_k)
\eeq
where $W_{i_1,i_2, \dots, i_k}^{(g)}(z_1,z_2, \dots, z_k)$ is the generating function of genus $g$ surfaces
with one boundary with the boundary condition induced by $W_{i_1,i_2, \dots, i_k}(z_1,z_2, \dots, z_k)$.
In this paper, we will be interested in the large $N$ limit of these correlation functions, that is to say, the generating
functions of discs with boundary operators of the form $W_{i_1,i_2, \dots, i_k}^{(0)}(z_1,z_2, \dots, z_k)$.

\subsection{Main result: disk amplitudes}

The main result of this paper is the computation of the mixed disk amplitudes with the use of a recursion relation explained in th.\ref{threc}:
\beq
{\cal W}_{1,j,j+1,\dots,\CN}(p_1,p_{j},\dots,p_\CN) = 
\Res_{q \to p_1,p_j^{+i,j}} {\cal K}_j(p_1,q_j,p_j) {\cal W}_{1,j+1,\dots, \CN}(q,p_{j+1},\dots, p_\CN)
\eeq
where the ingredients are explained in the following of the paper. In particular, this formula makes a heavy use of the different sheeted structures of the spectral curve of the model under study.

Another important feature of these notes are the explicit computation of these amplitudes as well as their expansion in the 3-matrix model. In section \ref{sec3matrix}, we apply the spectral curve approach as well as th.\ref{threc} to explicitly compute the number of tri-colored discs with given mixed boundary conditions.

\subsection{More notations}

In this section, we present the notations used in the following of these notes. We used the notations of
\cite{EPF} as long as it was appropriate so that the reader interested in both aspects of this matrix
model can go from one paper to the other without being confused.

First of all, let us remind the correlation functions to be computed:
\beq
W_{i_1,i_2, \dots, i_k}(x_1,x_2, \dots, x_k):= {T \over N} \left< \Tr {1 \over x_1 - M_{i_1}} {1 \over x_2 - M_{i_2}} \dots {1 \over x_k - M_{i_k}}  \right>.
\eeq

We also define the generating function of surfaces with two boundaries by
\beq
W_{\vec{i};\vec{j}}(x_1,x_2, \dots, x_k;y_1,\dots,y_l):=  \left< \Tr \left({1 \over x_1 - M_{i_1}} \dots {1 \over x_k - M_{i_k}}\right) \Tr \left({1 \over y_1 - M_{j_1}}\dots {1 \over y_l - M_{j_l}}\right) \right>_c
\eeq
where the index $c$ stands for the connected part and
\beq
\vec{i}  = (i_1,i_2,\dots,i_k)
\virg
\vec{j}  = (j_1,j_2,\dots,j_l)
\eeq
are sequences of integers in $[1,\CN]$.

Following \cite{Echain}, let us now define a set of polynomials $f_{i,j}(x_i, \dots, x_j)$ by induction:
\beq
\left\{\begin{array}{l}
f_{i,j} := 0 \qquad \hbox{if} \qquad j< i-1 \cr
f_{i,i-1}:=1 \cr
f_{i,i}(x_i) := {V_i'(x_i) \over c_{i,i+1}} \cr
c_{j,j+1} f_{i,j}(x_i,\dots,x_{j}):= V_{j}'(x_{j}) f_{i,j-1}(x_i,\dots,x_{j-1}) - c_{j-1,j} \, x_{j-1} \, x_{j} \, f_{i,j-2}(x_i, \dots, x_{j-2}) \cr
\end{array}
\right. .
\eeq

%\beq\begin{array}{rcl}
%f_{i,j}(x_i, \dots, x_j)&:=& {\displaystyle \prod_{k=i}^j} {1 \over c_{k-1,k}} \det\left|\begin{array}{cccc}
%V_i'(x_i) & -c_{i,i+1}x_{i+1} & & 0 \cr
%-c_{i,i+1}x_{i} & V_{i+1}'(x_{i+1}) & \ddots & \cr
% & \ddots & \ddots & - c_{j,j+1} x_j \cr
% 0 & & -c_{j,j+1} x_{j-1} & V_j'(x_j) \cr
% \end{array}\right| \;\; \hbox{if} \;\; i\leq j \cr
%&:=& 1 \;\; \hbox{if} \;\; i = j \cr
%&:=& 0 \;\; \hbox{if} \;\; i > j \cr
%\end{array}
%\eeq
%which are polynomials in all their variables. 

For $1\leq k \leq l \leq \CN$, let us then define 
\beq
P_{k,l;j_1, \dots, j_m}(x_1,\dots,x_\CN;y_1,\dots,y_m):=  \Pol_{x_k,\dots, x_l} f_{k,l}(x_k,\dots,x_l) W_{1,\dots,\CN;j_1,j_2, \dots, j_m}(x_1,\dots,x_\CN;y_1,\dots,y_m)
\eeq
which is a polynomial in the variables $x_i$, $i = k,\dots,l$.

%They allow to define the family of polynomials for $i=2,\dots,\CN-1$
%\beq
%P_i(x_1,x_i,x_{i+1}, \dots , x_\CN):= \Pol_{x_i,\dots, x_\CN} f_{i,\CN}(x_i,\dots, x_\CN)
%W_{1,i,i+1,\dots,\CN}(x_1,x_i,\dots,x_\CN)
%\eeq
%where $\Pol_{x_i,\dots, x_\CN} g(x_i,\dots,x_\CN)$ denotes the positive part of the Laurent expansion of
%$g(x_i,\dots,x_\CN)$ in all its variables, as well as the polynomial
%\beq
%P_1(x_1,x_2, \dots , x_\CN):= \Pol_{x_1,\dots, x_\CN} f_{1,\CN}(x_1,\dots, x_\CN)
%W_{1,2,\dots,\CN}(x_1\dots,x_\CN).
%\eeq

Given a pair of complex numbers $(x_1,x_2)$, we define a set of polynomial functions $\left\{\hat{x}_i(x_1,x_2)\right\}_{i=3}^\CN$ by the recursion relation
\beq\label{constrhat}
c_{i,i+1}\hat{x}_{i+1}:= V_i'(\hat{x}_i) - c_{i-1} \hat{x}_{i-1}
\eeq
for all $i=2, \dots,\CN$. The case $i=2$ is slightly different and we define
\beq
Y(x_1) := V_1'(x_1)-W_1(x_1).
\eeq
It should be noted that the latter function has a $1/N^2$ expansion.

Finally, it is convenient to denote
any observable whose variables satisfy the constraints \eq{constrhat} by:
\beq
\widehat{{\cal{O}}}(x_1,x_2):= {\cal{O}}(x_1,x_2,\hat{x}_3(x_1,x_2),\hat{x}_4(x_1,x_2),\dots,\hat{x}_\CN(x_1,x_2)).
\eeq

All the observables of the model admit a topological expansion \cite{thooft}, i.e. they can be written as a formal power
series in ${1 \over N^2}$:
\beq
{\cal{O}}= \sum_{g=0}^{\infty} N^{-2g} {\cal{O}}^{(g)}.
\eeq
%In this paper, we are interested only in the planar limit, that is to say the first term of this expansion, and we never consider the other terms of the expansion. We can thus forget about them in the following.
%

\section{Master loop equation and spectral curve}

Before going into the explicit computation of correlation functions, let us recall the algebraic structure underlying
the hermitian multi-matrix model studied by Eynard in \cite{Echain}. We will particularly focus on the sheeted structure
of the spectral curve since it is the most important ingredient needed in the following. The interested reader can
find a longer and more complete presentation of this structure in \cite{Echain} or more recently in \cite{EPF}.

\subsection{Loop equations}

The partition function and the correlation functions, seen as generating functions of random surfaces, are constrained by an infinite set of equations
prescribing how the weight of a surface is changed when one remove one edge from it. These equations where
first obtained by Tutte \cite{tutte,tutte2} for counting triangulated surfaces and later introduced under the name of loop
equation in the context of matrix models by Migdal \cite{Migdal}. These equations proved to be some very powerful tools since
they allowed the computation of many observables in different types of hermitian matrix models through the arising of algebraic
geometry.

Practically, these equations can be obtained by writing the invariance of the integral in \eq{defpartition}
under particular infinitesimal change of variables:
\beq
\begin{array}{l}
Z= \int dM_1 \dots dM_j \dots dM_{\CN} e^{- {N \over T} \Tr\left[{\displaystyle \sum_{k=1}^{\CN} V_k(M_k) - \sum_{k=2}^{\CN} c_{k-1,k} M_{k-1} M_k}\right] }
=\cr
\;\;\;\; =
\int dM_1 \dots d\tilde{M}_j \dots dM_{\CN} \, J(M_j,\tilde{M}_j) \, \times \cr
\qquad \quad e^{- {N \over T} \Tr\left[{\displaystyle \sum_{k\neq j} V_k(M_k) + V_j(\tilde{M}_j)- \sum_{k\neq j, j+1} c_{k-1,k} M_{k-1} M_k
 - c_{j-1,j} M_{j-1} \tilde{M}_j - c_{j,j+1} \tilde{M_j} M_{j+1} }\right] } \cr
\end{array}
\eeq
where $\tilde{M}_j := M_j + \epsilon \, \delta M_j$ is an infinitesimal deformation of $M_j$. The loop equation
just corresponds to saying that the variation of the action is compensated by the Jacobian of this change of variable to first order in
$\epsilon$.

The set of loop equations corresponding to the changes of variables
\beq
M_k \to M_k + \epsilon \, {1\over x_{k+1}-M_{k+1}} {1\over x_{k+2}-M_{k+2}} \dots {1\over x_{\CN}-M_{\CN}} {1\over x_{1}-M_{1}}
\eeq
supplemented with some algebra allows to show that
\bt
For $0<k<\CN$, the correlation functions satisfy
\bea\label{masterloop1}
\left(x_k - \hat{x}_k(x_1)\right) W_{1,k,\dots,\CN}(x_1,x_k,\dots,x_\CN) &=& W_{1,k+1,\dots,\CN}(x_1,x_{k+1},\dots,x_\CN) \cr
&& - P_{1,k-1}(x_1,\hat{x}_2,\dots,\hat{x}_{k-1},x_k,\dots,x_\CN)\cr
&& + {T \over N^2} P_{2,k-1;1}(x_1,\hat{x}_2,\dots,\hat{x}_{k-1},x_k,\dots,x_\CN)\cr
\eea
where 
\beq
P_{i,k}(x_1,\dots,x_\CN):= \Pol_{x_i,\dots,x_{k}}f_{i,k}(x_1,\dots,x_{k-1}) W_{1,\dots,\CN}(x_1,\dots,x_\CN)
\eeq
is a polynomial in the variables $x_j$, $j = i, \dots, k$ and the polynomials $f_{i,k}(x_i,\dots,x_k)$ are defined by induction through
\beq
 f_{i,k}(x_i,\dots,x_{k-1}) = {V_k'(x_k) f_{i,k-1} - c_{k-1,k} x_{k-1} x_k f_{i,k-2}\over c_{k,k+1}}.
 \eeq

For $k = \CN$, one has
\beq\label{masterloopN}
\left(V_\CN'(\hat{x}_\CN) - \hat{x}_{\CN-1}\right) W_{1}(x_1) = P_{1,\CN}(x_1,\hat{x}_2,\dots,\hat{x}_\CN) + {T^2 \over N^2}  P_{2,\CN;1}(x_1,\hat{x}_2,\dots,\hat{x}_\CN;x_1) .
\eeq

\et 

This theorem was proved in \cite{Echain,EPF}. For completeness, it is  derived for the present model with the notations in hand in appendix A.

\subsection{Leading order and master loop equation}

By plugging $Y(x) = V_1'(x) - T W_1(x)$ into \eq{masterloop1}, the leading order in the $1/N^2$ expansion of this equation reads
\beq\label{masterloop}
\hat{E}(x_1,y(x_1)) = 0.
\eeq
where
\beq
y(x) = Y^{(0)}(x)
\eeq
is the leading order of the ${1 \over N^2}$ expansion of $Y(x)$
and
\beq
E(x_1,x_2, \dots, x_\CN) = (V_1'(x_1) - x_2)(V_\CN(x_\CN) - x_{\CN-1}) - P_{1,\CN}^{(0)}(x_1, \dots, x_\CN)
\eeq
is a polynomial in all its variables.

%
%Let us define
%
%
%By carefully choosing the change of variables, one can obtain a hierarchy of equations fixing unambiguously the value of a large family of correlation functions. This set of loop equations implies \cite{Echain,EPF}, to leading order in the topological expansion, the so-called
%{\em master loop equation}:
%\beq\label{masterloop}\encadremath{
%\hat{E}(x_1,x_2)= \left(c_{1,2}x_2-y(x_1)\right) \hat{P}_2(x_1,x_2).}
%\eeq
%where
%\beq
%E(x_1,x_2, \dots, x_\CN) = (V_1'(x_1) - x_2)(V_\CN(x_\CN) - x_{\CN-1}) - P_{1,\CN}^{(0)}(x_1, \dots, x_\CN)
%\eeq
%is a polynomial in all its variables.
%
%For $x_2=y(x_1)$, it reduces to the algebraic equation
%\beq\label{masterloop2}
%\hat{E}(x_1,y(x_1)) = 0.
%\eeq

Consider \eq{masterloop} as an equation with unknown $x_1$ and $y(x_1)$. On can see that this is a polynomial equation
of degree ${\displaystyle \prod_{i = 1}^{\CN}} d_i$ in $x_1$ and $1 + {\displaystyle \prod_{i = 2}^{\CN}} d_i$ in $y(x_1)$. Thus, it defines an algebraic
curve ${\cal E}(x,y)=0$ in $C^2$.

\subsection{Description of the spectral curve}

Let us now consider the algebraic equation
\beq
\hat{E}(x,y) = 0.
\eeq
It defines an algebraic curve in $\mathbb{C}^2$ which encodes most of the combinatorial properties of the matrix model: it is the 
spectral curve associated to the model.

In this section, we study the properties the algebraic curve $\hat{E}(x,y) = 0$ as a compact Riemann surface ${\cal{L}}$ equipped with
two meromorphic functions $x$ and $y$ such that
\beq
\forall \, p \in {\cal L} \, , \; \, \hat{E}(x(p),y(p)) = 0.
\eeq
A point $p\in {\cal L}$ is equivalent to a pair $(x(p),y(p))\in \mathbb{C}^2$ satisfying $\hat{E}(x(p),y(p)) = 0$.

Since, given a pair $(x,y)\in \mathbb{C}^2$, all the $\hat{x}_i(x,y)$ are unambiguously fixed, one can see them as monovalued meromorphic functions
$z_i$ on ${\cal{L}}$ such that:
%\footnote{As we will see later, one can find more than one point $p$ on
%${\cal{L}}$ corresponding to one fixed value of $x(p)$. Nevertheless, this ambiguity correspond to the multivaluedness of
%the function $z_i(p)$ and we present later on the procedure leading to the physical value.}:
\beq
\forall p \in {\cal{L}} \, , \; \; z_i(p) = \hat{x}_i(x(p),y(p)).
\eeq
By definition, one has
\beq
x(p)  = z_1(p) \virg y(p) = z_2(p).
\eeq

Using these functions, one can write:
\beq
\forall p \in {\cal{L}} \; , \; \; E(z_1(p), z_2(p), z_3(p),  \dots , z_\CN(p)) = 0
\eeq
and one can choose to fix any of the functions $z_i(p)$ as parameter to study the curve. Depending on which
parameter we chose we obtain different descriptions that we can now characterize.

\subsubsection{Sheeted structure and points at infinity}

Let us first study this curve in terms of the ``physical" variable $x:=z_1$. Since $\hat{E}(x,y)$ is a polynomial
in $y$ of degree $1 + {\displaystyle \prod_{i = 2}^{\CN}} d_i$, $y(x)$ is a $1 + {\displaystyle \prod_{i = 2}^{\CN}} d_i$-valued
functions of $x$. This means that, for a generic value of $x$, there exist $1 + {\displaystyle \prod_{i = 2}^{\CN}} d_i$
points $p^j$ on ${\cal{L}}$  such that:
\beq
\forall i = 0, \dots, {\displaystyle \prod_{i = 2}^{\CN}} d_i \; , \;\; z_1(p^i) = z_1(p).
\eeq
Hence ${\cal{L}}$ can be seen as $1+s_1$ copies of the Riemann sphere, called $z_1$-sheets, glued together by cuts
in such a way that $x(p) = z_1(p)$ is injective in each sheet. This sheeted structure reflects the multi-valuedness of
the resolvent $W_1^{(0)}(x)$ as a function of $x$. Indeed, once the value of this correlation function is fixed,
$y(p) = z_2(p)$ is fixed and
all the $z_i(p) = \hat{x}_i(x(p),y(p))$'s are determined. Thus, this fixes one unique point in ${\cal{L}}$.

How can we extract the right value of $y(x)$ giving access to the generating functions of random surfaces?
In order to distinguish the different $x$-sheets, one can look at the pre-images of $\infty$ by $x(p)$, i.e.
the points $\infty^j \in {\cal{L}}$ such that $x(\infty^j)= \infty$.

From its definition as an asymptotic series, one knows that the physical solution of the algebraic equation should satisfy\footnote{We call this solution physical because it is the branch whose expansion in $x$ gives rise to the generating function of surfaces.}:
\beq
y(p) \sim_{x(p) \to \infty}  V'(x(p)) + {1 \over x(p)} + O(x^{-2}(p)).
\eeq

It means that there exists at least one pre-image of infinity, denoted $\infty^1 \in {\cal L}$, such that
\beq
z_2(p) \sim_{p \to \infty^1} V'_1(z_1(p)) + {1 \over z_1(p)} + O(z_1^{-2}(p))
\eeq
and more generally
\beq
\forall k = 2, \dots , \CN \; , \; \; z_k(p) \sim_{p \to \infty^1} O(z_1^{s_k}) 
\eeq
where
\beq
s_k = {\displaystyle \prod{i=1}^{k-1}} d_i
\eeq
One could symmetrically choose $z_\CN$ as a variable and use the physical condition that
\beq
z_{\CN-1}(p)  \sim_{z_\CN(p) \to \infty} V'_\CN(z_{\CN}(p)) + {1 \over z_\CN(p)} + O(z_\CN^{-2}(p)).
\eeq
This implies that there exist at least one pre-image $\infty^\CN$ of infinity by $x$ such that:
\beq
z_1(p) \sim_{p \to \infty^{\CN}} O(z_\CN(p)^{r_1})
\eeq
where
\beq
r_k = {\displaystyle \prod{i=k+1}^{\CN}} d_i.
\eeq
This means that $r_1$ different $z_1$-sheets merge at $\infty^\CN$.
Since $r_1+1$ is exactly the number of $z_1$-sheets, there exist no other pre-image of infinity by $x = z_1$ and the fiber above infinity reads: $x^{-1}(\infty) = \left\{\infty^1,\infty^\CN\right\}$.

\begin{figure}
  \begin{center}
  \includegraphics[width=5cm,angle=00]{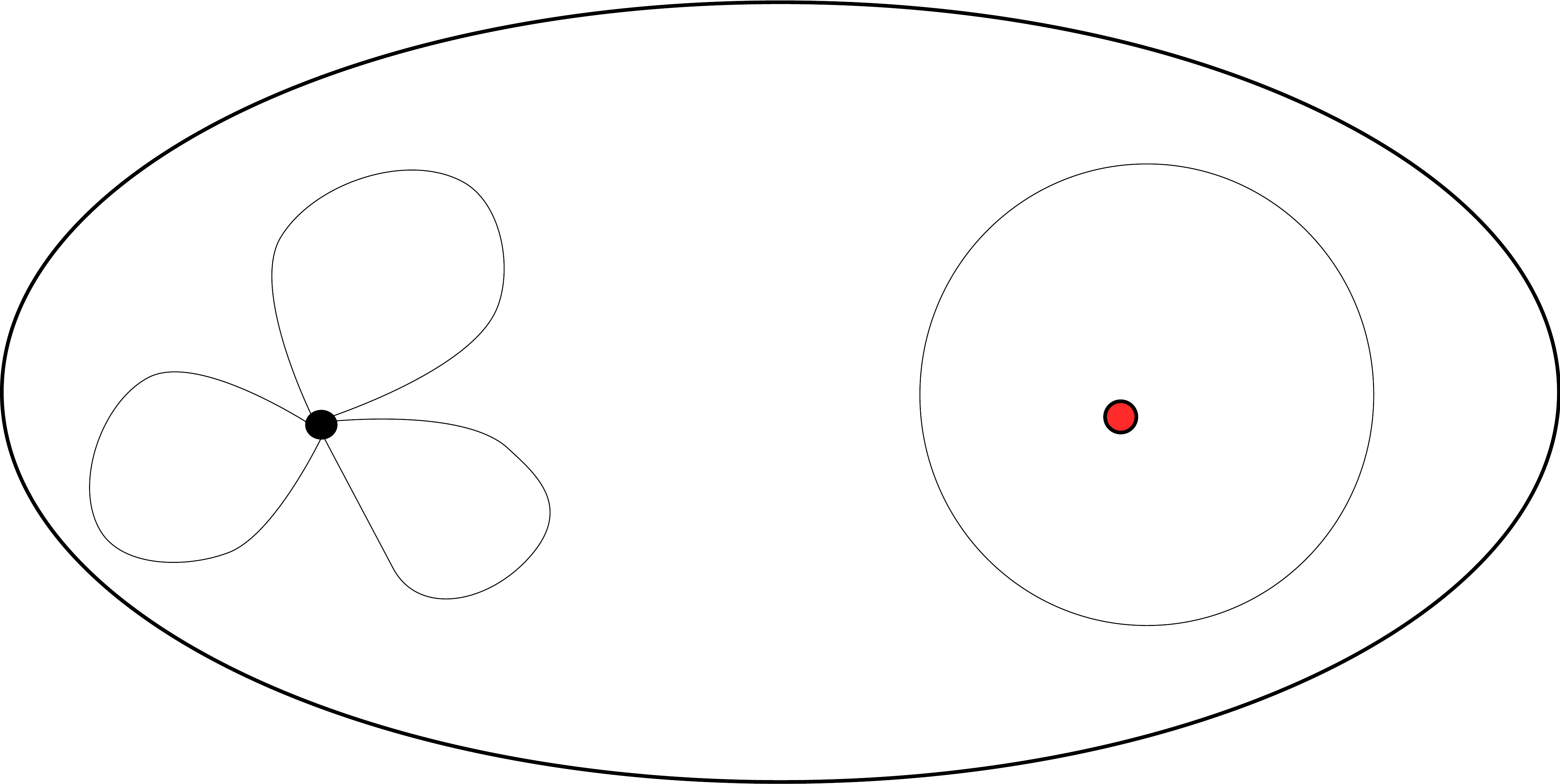}\\
  \end{center}
  \caption{ Example of sheeted structure of ${\cal L}$ where one has represented a genus 0 spectral curve as the Riemann sphere, the red dot represents the pole $\infty^1$ while the black dot is the other pole $\infty^\CN$. The different sheets are represented by the connected components separated by continuous lines. In this case, $r_1 = 4$ of them join at $\infty^\CN$ and only one contains $\infty^1$.}\label{sheetedfig}
\end{figure}

One can summarize the structure of $\hat{E}$ in terms of $z_1$ as follows\footnote{An example is depicted in fig.\ref{sheetedfig}.}:
\begin{itemize}
\item ${\cal{L}}$ is composed of $1+r_1$ $z_1$-sheets;

\item $z_1(p)$ has two poles in ${\cal{L}}$: one simple pole noted $\infty^1$
and one pole of degree $r_1$ noted $\infty^\CN$;

\item $\infty^1$ belongs to only one sheet called the physical sheet since it corresponds to physical solutions. All other
sheets merge at $\infty^\CN$.
\end{itemize}

%
%One can represent this description of ${\cal{L}}$ by:
%\beq
%\begin{array}{r}
%{\epsfysize 4cm\epsffile{sheetedspcurve.eps}}
%\end{array}
%\eeq
%where one has represented a genus 0 spectral curve as the Riemann sphere, the red dot represents the pole $\infty^1$ while the black dot is the other pole $\infty^\CN$. The different sheets are represented by the connected components separated by continuous lines. In this case, $r_1 = 4$ of them join at $\infty^\CN$ and only one contains $\infty^1$.

The unique $z_1$-sheet containing $\infty^1$ is only referred to as the physical sheet. Indeed, in order to recover the physically meaningful quantities $\Tr M_1^k$, one has to expand $y(z)$ around $z \to \infty^1$:
\beq
y(z) \sim_{z \to \infty^1} V_1'(x(z)) + \sum_{k = 0}^\infty {1 \over x(z)^{k+1}} \, \lim_{N \to \infty}{1 \over N} \left<\Tr M_1^k\right>.
\eeq
The expansion of $y(z)$ around $\infty^\CN$ doesn't have such a simple combinatorial interpretation.

One can also fix $z_k(p)$ and look at the $z_k$-sheeted structure of ${\cal{L}}$. With the same arguments, one can see that:
\begin{itemize}
\item There exists $r_k+s_k$ $z_k$-sheets where $z_k(p)$ is injective;

\item $z_k(p)$ has two poles. One pole of degree $s_k$ in $\infty^1$ where $s_k$ $z_k$-sheets meet and one
pole of degree $r_k$ in $\infty^\CN$ where the other $r_k$ $z_k$-sheets merge.

\item  One notes $p^{+j,k}$ (resp. $p^{-j,k}$) for $j = 1, \dots , s_k$ (resp. $j = 1, \dots , r_k$) the pre-images of $z_k(p)$ belonging to the different sheets
which merge at $\infty^1$ (resp. $\infty^\CN$).

\end{itemize}

In particular, one can see that there exists a neighborhood of $\infty^1$ where $z_1$ is injective in the $z_k$-sheets.
This means that:
\beq\label{injectivite}
z_k(p^{+i,k}) = z_k(q^{+j,k}) \Rightarrow x(p^{+i,k}) = x(q^{+j,k})
\eeq
in some neighborhood of $\infty^1$.

%\subsubsection{Holomorphic differentials and cycles}
%
%From the properties of the polynomial $\widehat{E}(x,y)$, one knows that the Riemann surface ${\cal L}$ has genus ${\cal G} \leq \prod_k d_k -1$. Let us $\left\{{\cal A}_i,{\cal B}_i\right\}_{i = 1}^{\cal G}$ be a canonical basis of cycles on ${\cal L}$.
%
%Let $\left\{du_i\right\}_{i = 1}^{\cal G}$ be the dual basis of holomorphic one-forms normalized on the ${\cal A}$ cycles by
%\beq
%\oint_{p\in {\cal A}_j} du_i(p) = \delta_{i,j}.
%\eeq

%\subsubsection{Bergman Kernel}
%
%There exist a unique one-form  $B(p,q)$ on ${\cal L} \times {\cal L}$, the Bergman kernel, with the following properties:
%\begin{itemize}
%
%\item As a differential in its first variable $p$, it has a unique pole located at $p = q$ with the behavior
%\beq
%B(p,q) \sim {dp \, dq \over (p-q)^2} + \; \hbox{regular}
%\eeq
%in local coordinates.
%
%\item Its periods along the ${\cal A}$ cycles are normalized
%\beq
%\oint_{p \in {\cal A}_i} B(p,q) = 0
%\eeq
%for $i = 1, \dots, {\cal G}$.
%
%\end{itemize}

\section{Correlation functions}

\subsection{State of the art}

Let us first recall the results of \cite{Echain,EPF}.

In \cite{EPF}, a very efficient inductive method has been found for computing the whole topological expansion of the free energy as well as the correlation functions involving only the first matrix of the chain. This means that the free energy as well as the correlation functions of the type $\left<{\displaystyle \prod_{i=1}^k} \Tr {1 \over x_k - M_1}\right>$ are known to any order in the large $N$ expansion.

In \cite{Echain}, some other observable were computed. First of all, the leading order in the large $N$ expansion of any two point correlation function of the type $\left< \Tr{1 \over x_1-M_i} \Tr {1 \over x_2-M_j}\right>_c$ was computed and expressed in terms of a fundamental one form on ${\cal L} \times {\cal L}$ often called the Bergman kernel. More important to us is the derivation of the simplest disk amplitude with mixed boundary conditions, that is the large $N$ limit of the correlation function $\left<\Tr {1 \over x_1-M_1} {1 \over x_2 - M_{\cal N}}\right>$.

\subsection{Mixed correlation functions}

With this description of the algebraic curve in hand, we are ready to compute the correlation functions.
Let us first promote them to mono-valued differentials ${\cal W}_{i_1, \dots i_k}(p_1, \dots, p_k)$ on ${\cal L}^k$
defined by:
\beq
{\cal W}_{i_1, \dots i_k}(p_1, \dots, p_k) = W_{i_1, \dots i_k}(z_{i_1}(p_1), \dots, z_{i_k}(p_k)) \, dz_{i_1}(p_1), \dots, dz_{i_k}(p_k).
\eeq
We use the same notation for any other function considered in the previous section: the curly letters represent differentials on the spectral curve.

From now on, we will consider only the leading order of the correlation functions in the $1/N^2$ expansion. We thus  abusively denote
\beq
W_{i_1,i_2, \dots, i_k}(x_1,x_2, \dots, x_k):=W_{i_1,i_2, \dots, i_k}^{(0)}(x_1,x_2, \dots, x_k)
\eeq
and
\beq
P_{k,l}(x_1, \dots , x_\CN):=P_{k,l}^{(0)}(x_1, \dots , x_\CN)
\eeq
the leading orders of the observables we are studying. Let us also remind that
\beq
y(x_1):=Y^{(0)}(x_1) = V_1'(x_1)-W_1^{(0)}(x_1).
\eeq

%To leading order in the $1/N^2$ expansion, the loop equation \eq{masterloop1} reads
%\bea \label{loop1}
%\left(x_k - \hat{x}_k(x_1)\right) W_{1,k,\dots,\CN}(x_1,x_k,\dots,x_\CN) &=& W_{1,k+1,\dots,\CN}(x_1,x_{k+1},\dots,x_\CN) \cr
%&& - P_{1,k-1}(x_1,\hat{x}_2,\dots,\hat{x}_{k-1},x_k,\dots,x_\CN).\cr
%\eea
In terms of  differentials on the spectral curve, to leading order in the $1/N^2$ expansion, the loop equation \eq{masterloop1} can then be written:
\bea\label{loop2}
(z_j(p_j)-z_j(p_1)) {\cal W}_{1,j,j+1, \dots , {\cal N}}(p_1, p_j, \dots , p_{\cal N}) &=& {\cal W}_{1,j+1, \dots , \CN}(p_1,p_{j+1}, \dots, p_{\CN}) dz_j(p_j)\cr
&& - {\cal P}_{1,j-1}(p_1,z_2(p_1), \dots , z_{j-1}(p_1), p_j, \dots, p_\CN)\cr
\eea
where
\beq
{{\cal P}_{1,j-1}(p_1,z_2(p_1), \dots , z_{j-1}(p_1), p_j, \dots, p_\CN) \over {\displaystyle \prod_{i=j}^{\CN}} dz_i(p_i)}  = P_{1,j-1}(z_1(p_1), \dots , z_{j-1}(p_1), z_j(p_j), \dots, z_\CN(p_\CN)) 
\eeq
is a polynomial in $z_1(p_1)$ of degree $r_{j}-1$.

The main result of these notes is that one can solve this equation by induction on $j$. Indeed, \eq{loop2} expresses  ${\cal W}_{1,j,j+1, \dots , {\cal N}}(p_1, p_j, \dots , p_{\cal N})$ in terms of ${\cal W}_{1,j+1, \dots , {\cal N}}(p_1, p_j, \dots , p_{\cal N})$ provided that one knows the polynomial ${\cal P}_{1,j-1}(p_1,z_2(p_1), \dots , z_{j-1}(p_1), p_j, \dots, p_\CN)$. The polynomiality of the later actually allows one to compute it by using only \eq{loop2} and get 
\bt\label{threc}
{\bf Recursion relation for the disk amplitudes.}

The mixed disk amplitudes satisfy the recursion relation

\beq
{\cal W}_{1,j,j+1,\dots,\CN}(p_1,p_{j},\dots,p_\CN)= \Res_{q \to p_1,p_j^{+i,j}}
{{\cal W}_{1,j+1,\dots,\CN}(q,p_{j+1},\dots,p_\CN) {\displaystyle \prod_{k=1}^{s_{j}}}\left[z_{1}(p_1)-z_{1}(p_j^{+k,j})\right]\over
\left[z_{1}(q)-z_{1}(p_1)\right] \left[z_{j}(p_j)-z_j(p_1)\right]  {\displaystyle \prod_{k=1}^{s_{j}}}
\left[z_{1}(q)-z_{1}(p_j^{+k,j})\right]}.
\eeq

\et

Since the initial value
\beq
{\cal W}_{1,\CN}(p_1,p_\CN) = \sum_{k=1}^\CN\frac{E\left(Z_1,...,Z_k,\tilde{Z}_{k+1},...,\tilde{Z}_{\mathcal{N}}\right)}{\left(Z_k-\tilde{Z}_k\right)\left(Z_{k+1}-\tilde{Z}_{k+1}\right)}
\eeq
\beq
Z_k=z_k\left(p_1^{+,1}(z_1)\right)\:\: \tilde{Z}_k=z_k\left(p_\CN^{-,\CN}(z_\CN)\right)
\eeq
was computed in \cite{Echain}, this theorem allows to compute ${\cal W}_{1,j,j+1,\dots,\CN}(p_1,p_{j},\dots,p_\CN)$ for any $j=2,\dots,\CN-1$ and in particular the complete mixed correlation function:
\bc 
\label{correc}
The disk amplitudes read, for $j=2,\dots, \CN-1$:
\beq
{\cal W}_{1,j,j+1,\dots,\CN}(p_1,p_{j},\dots,p_\CN) = 
\Res_{q_j} \Res_{q_{j+1} } \dots \Res_{q_{\CN-1} } \prod_{\alpha =j}^{\CN-1} {\cal K}_\alpha(p_1,q_\alpha,p_\alpha) {\cal W}_{1,\CN}(q_{\CN-1},p_\CN)
\eeq
where one defines the recursion kernel
\beq
{\cal K}_\alpha(p,q,r)  = {{\displaystyle \prod_{k=1}^{s_{\alpha}}}\left[z_{1}(p)-z_{1}(r^{+k,\alpha})\right]\over
\left[z_{1}(q)-z_{1}(p)\right] \left[z_{\alpha}(r)-z_\alpha(p)\right]  {\displaystyle \prod_{k=1}^{s_{\alpha}}}
\left[z_{1}(q)-z_{1}(r^{+k,\alpha})\right]}
\eeq
and the residue sign stands for
\beq
\forall i = j, \dots , \CN-1 \, , \; \Res_{q_i} = \Res_{q_i \to p_1} + \sum_{k=1}^{s_i} \Res_{q_i \to p_{i}^{+k,i}}.
\eeq

\ec

In turn, this gives access to any disk amplitude of the form ${\cal W}_{i_1,\dots,i_k}$ with $1\leq i_1<i_2<\dots<i_k\leq i_\CN$ by considering the leading order of the expansion the complete disk amplitude as some $p_i \to \infty^1$.

\bigskip

\subsection{Proof of the recursion relation}

For proving th.\ref{threc}, one first needs a simple technical lemma.
\bl

$(z_j(p_j)-z_j(p_1)) {\cal W}_{1,j,j+1, \dots , {\cal N}}(p_1, p_j, \dots , p_{\cal N})$ vanishes when $p_1 = p_j^{+i,j}$, for $i=1, \dots, s_k$.
\el

\proof{
The proof follows from the combinatorial interpretation of $ {\cal W}_{1,j,j+1, \dots , {\cal N}}(p_1, p_j, \dots , p_{\cal N})$ when $p_1$ lies in the physical sheet.

Indeed, for $p_1 \to \infty^1$ (and not in any other patch of the spectral curve), the definition of ${\cal W}_{1,j,j+1, \dots , {\cal N}}(p_1, p_j, \dots , p_{\cal N})$ reads
\bea
&& N \, (z_j(p_j)-z_j(p_1)) \, {\cal W}_{1,j,j+1, \dots , {\cal N}}(p_1, p_j, \dots , p_{\cal N})  =  \cr
&& \qquad \qquad = (z_j(p_j)-z_j(p_1)) \left< \Tr {1 \over z_1(p_1) -M_1} {1 \over z_j(p_j) -M_j} \dots {1 \over z_\CN(p_\CN) -M_\CN} \right>.
\eea
One can thus write
\bea
&& N (z_j(p_j)-z_j(p_1)) {\cal W}_{1,j,j+1, \dots , {\cal N}}(p_1, p_j, \dots , p_{\cal N}) \cr
&=& \left< \Tr {1 \over z_1(p_1) -M_1} {1 \over z_{j+1}(p_{j+1}) -M_{j+1}} \dots {1 \over z_\CN(p_\CN) -M_\CN} \right> \cr
&&  - \left< \Tr {1 \over z_1(p_1) -M_1} {z_j(p_1) - M_j \over z_j(p_j) -M_j} \dots {1 \over z_\CN(p_\CN) -M_\CN} \right>\cr
\eea
which vanishes when $z_j(p_j) = z_j(p_1)$, i.e. $p_1$ belongs to the same $z_j$-fiber as $p_j$. Since this formula is valid only for $p_1$ in the $z_1$ physical sheet, this additional constraints implies that $(z_j(p_j)-z_j(p_1)) {\cal W}_{1,j,j+1, \dots , {\cal N}}(p_1, p_j, \dots , p_{\cal N})$ vanishes for $p_1 = p_j^{+i,j}$.}

\bigskip

{\bf Proof of theorem \ref{threc}}

Thus, one knows that the left hand side of \eq{loop2}, vanishes for $p_1 = p_j^{+i,j}$.

On the other hand, the second term of the right hand side,
$P_{1,j-1}$, is a polynomial in
$z_1(p_1)$ of degree $s_j-1$. Thanks to the vanishing of the left hand side,  one knows its
value in the $d_1 \dots d_{j-1}$ points, $z_1(p_j^{+i,j})$
for $j=1, \dots, s_j$ :
\beq P_{1,j-1}(z_1(p_j^{+i,j}),z_2(p_j^{+i,j}),\dots
,z_{j-1}(p_j^{+i,j}),z_{j}(p_j)\dots,,z_\CN(p_{\CN}))
=W_{1,j+1,\dots,\CN}(p_j^{+i,j},p_{j+1},\dots,p_\CN). \eeq

By Lagrange interpolation, one gets an explicit expression for this polynomial which one can plug back into \eq{masterloop1} to get
\bea 
&&(z_{j}(p_j)-z_j(p_1))
{\cal W}_{1,j,\dots,\CN}(p_1,p_{j},\dots,p_\CN)= {\cal W}_{1,j+1,\dots,\CN}(p_1,p_{j+1},\dots,p_\CN) \cr &&-
 \sum_{i=1}^{s_j}
{{\cal W}_{1,j+1,\dots,\CN}(p_j^{+i,j},p_{j+1},\dots,p_\CN) {\displaystyle \prod_{k=1}^{s_j}}\left[z_{1}(p_1)-z_{1}(p_j^{+k,j})\right]\over
\left[z_{1}(p_1)-z_{1}(p_j^{+i,j})\right] {\displaystyle \prod_{k\neq i}}
\left[z_{1}(p_j^{+i,j})-z_{1}(p_j^{+k,j})\right]}.
 \eea

This last equation can be written under the form of a residue formula
\beq
{\cal W}_{1,j,\dots,\CN}(p_1,p_{j},\dots,p_\CN)= \Res_{q \to p_1,p_j^{+i,j}}
{{\cal W}_{1,j+1,\dots,\CN}(q,p_{j+1},\dots,p_\CN) {\displaystyle \prod_{k=1}^{s_j}}\left[z_{1}(p_1)-z_{1}(p_j^{+k,j})\right]\over
\left[z_{1}(q)-z_{1}(p_1)\right] \left[z_{j}(p_j)-z_j(p_1)\right]  {\displaystyle \prod_{k=1}^{s_j}}
\left[z_{1}(q)-z_{1}(p_j^{+k,j})\right]}.
\eeq
$\square$

\section{Example of application: the 3-matrix model}

\label{sec3matrix}

In this section, we perform explicit computations in the case $\CN = 3$ resulting in the computation of three colored discs with mixed boundary conditions.

%\subsection{Definitions:}
%\beq
%W_i(z) = {1 \over N} \moy{\Tr {1 \over z-M_i}}
%\eeq
%
%\beq
%Z_1(z) = W_0(z) - V_0'(z)
%\eeq
%
%\bea
%U_i(z) &=& Pol_z V_i'(z) W_i(z) \cr
%&=& {1 \over N} \moy{\Tr {V_i'(z)-V_i'(M_i) \over z-M_i}} \cr
%\eea

\subsection{Genus 0 spectral curve}

For enumerating maps, the spectral curve has to be a rational curve, i.e. its genus has to be vanishing. Hence, there  exists a global coordinate $p$ such that the functions $z_i$ are rational functions \cite{Emap}
\beq
\forall i = 1,\dots,3 \, , \; z_i(p) = \sum_{k= -s_i}^{r_i} \alpha_{i,k}\: p^k
\eeq
satisfying

\beq\label{condi}
V'_1(z_1(p))-z_2(p) \sim_{p \to \infty} {T \over p\:\alpha_{1,1} }  + O\left(p^{-2}\right)
\eeq
\beq
V'_3(z_3(p))-z_2(p) \sim_{p \to 0} {Tp \over \alpha_{3,-1} }  + O\left(p^{2}\right)
\eeq
\beq
V'_2(z_2(p))=z_1(p)+z_3(p),
\eeq
where $\alpha_{1,1}=\alpha_{3,-1}=\gamma$. The limits of the sums describing the parametrizations are obtained from the degrees of the potentials, and are given by  $s_1=r_3=1$,  $r_1=d_2d_3$, $s_2=d_1$,  $r_2=d_3$ and $s_3=d_1d_2$.  The solution branch giving rise to the spectral curve is the one where the coefficients $\gamma$ and $\alpha_{i,k}$ are algebraic functions of $T$ such that $\gamma \to 0$ as $T \to 0$.
With this global parameterization $\infty_1$ is located in $p = \infty$ whereas $\infty_3$ is $p = 0$.

Once the spectral curve, i.e. the collection of rational functions $z_i(p)$, is known, one can study its different sheeted structures.
The $z_i$-fiber over a point $Z_i$ is given by the set of solution $p^{\pm j,i}(Z_i)$ satisfying
\beq
z_i\left(p^{\pm j,i}(Z_i)\right) = Z_i.
\eeq
Among these solutions, $p^{+j,i}(Z)\to \infty$ and  $p^{-j,i}(Z)\to 0$ as $Z \to \infty$.

\br
In the setup of enumerative geometry, the exponent of $T$ is the generating functions is the number of vertices in the maps considered. Thus, as long as one is interested in sufficiently small maps, the preceding equations need being solved only up to a given order in the $T$ expansion.

One can thus compute the fibers above a point $Z$ through its small $T$ expansion.

\er

\subsection{The complete one loop function}

\subsubsection{Loop equations and analytic solution}

Since the 3-matrix model is the simplest model which carries the whole complexity of the chain of matrices, let us use it as a toy model for deriving more carefully the loop equations involved in the computation of the complete mixed correlation function.

Let us consider the following change of variables 
\beq
M_1\rightarrow M_1+\epsilon \delta M_1
\eeq
where
\beq
\delta M_1=\frac{1}{z_2-M_2}\frac{1}{z_3-M_3}\frac{1}{z_1-M_1}.
\eeq
In order to keep the path-integral invariant under such a change, we demand that
the contributions from the Jacobian of the change of variables and the contributions from
the variation of the action cancel each other. To order $\epsilon$, we have
\beq
\moy{Tr\left(\frac{\partial \delta M_1}{\partial M_1}\right)}=-N\left(\moy{Tr(V'_1(M_1)\delta M_1)}-\moy{Tr(\delta M_1M_2)}\right)
\eeq
The left hand side can be written in terms of the connected correlators
\beq
\moy{Tr(\delta M_1)Tr\left(\frac{1}{z_1-M_1}\right)}=\moy{Tr(\delta M_1)}_c\moy{Tr\left(\frac{1}{z_1-M_1}\right)}_c+\moy{Tr(\delta M_1)Tr\left(\frac{1}{z_1-M_1}\right)}_c,
\eeq
therefore we can write the left hand side in terms of correlators dependent on the points $p,q,r\in\Sigma$ as
\beq
N^2{\cal W}_{1,2,3}(p,q,r){\cal W}_{1}(p)+{\cal W}_{1,2,3;1}(p,q,r;p)
\eeq
where
\beq
z_1(p) = z_1\virg z_2(q) = z_2 \virg z_3(r) = z_3.
\eeq

On the other hand the right hand side can be expressed as
\beq
-N\moy{Tr((V'_1(M_1)-V'_1(z_1))\delta M_1)}+N^2(V'(z_1(p))+z_2(q)){\cal W}_{1,2,3}(p,q,r)-N^2{\cal W}_{1,3}(p,r).
\eeq
By collecting the terms and considering only the order $N^0$ terms, we have 
\beq
(z_2(q)-z_2(p)) {\cal W}_{1,2,3}(p,q,r) = {\cal W}_{1,3}(p,r) - \Pol_{z_1(p)} V_1'(z_1(p)) {\cal W}_{1,2,3}(p,q,r),
\eeq
where
\beq
{\cal W}_{1,3}(p,r) =\frac{1}{Z_2-\tilde{Z}_2}\left(\frac{E(z_1,\tilde{Z}_2,z_3)}{z_1-\tilde{Z}_1}-\frac{E(z_1,Z_2,z_3)}{z_3-Z_3}\right)
\eeq
is known from \cite{Echain}, and where we used the notation: $Z_2=V'_1(z_1)-W_1(z_1)$, $\tilde{Z}_2=V'_3(z_3)-W_3(z_3)$, $Z_3=V'_2(Z_2)-z_1$
and $\tilde{Z}_1=V'_2(\tilde{Z}_2)-z_3$.

This equation can be solved and its solution is given by th.\ref{threc} under the form
\beq\label{3mm}
{\cal W}_{1,2,3}(p_1,p_2,p_3) =-Res_{Q\rightarrow p_1,p_3^{+i,3}}\left( \frac{{\cal W}_{1,3}(Q,p_3){\displaystyle\prod_{k=1}^{s_2}(z_1(p_1)-z_1(p_2^{+k,2}))}}{(z_1(Q)-z_1(p_1))(z_3(p_3)-z_3(p_1))){\displaystyle\prod_{k=1}^{s_2}(z_1(Q)-z_1(p_2^{+k,2}))}}\right).
\eeq

\subsubsection{Triangulations}

In this section we present some explicit computations for the three matrix model. We assume that all of the matrices are subject to cubic potentials of the form
\beq
 V_i(x)=\frac{1}{2}g_2^{(i)}x^2+\frac{1}{3}g_3^{(i)}x^3.
 \eeq

Furthermore, for simplicity we assume $g_2^{(1)}=g_2^{(3)}=1$ and $g_2^{(2)}=3$ in order to obtain a convenient form for the propagator matrix:

\beq
C^{-1} :=\left(
\begin{array}{ccc}
2&-1&1\cr
-1&1&-1\cr
1&-1&2\cr
\end{array}
\right) .
\eeq

\begin{figure}
  \begin{center}
  \includegraphics[width=12cm,angle=00]{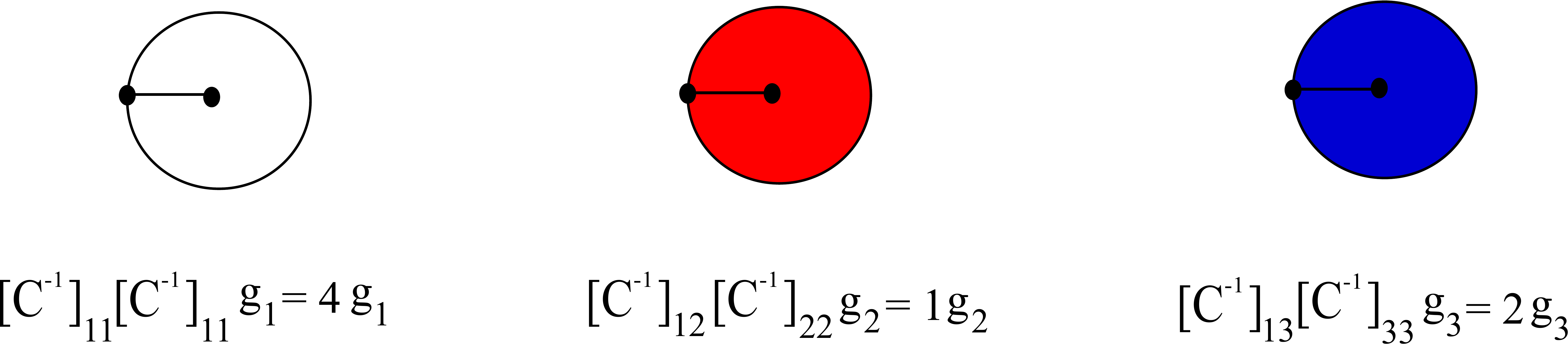}\\
  \end{center}
  \caption{ Maps contributing to ${\cal T}_{1,0,0}^2$. The type 1 (resp. 2 and 3) boundary condition is represented by the color white (resp. red and blue). The outer face is the marked face of type 1.}\label{2vertex}
\end{figure}

The conditions \ref{condi}, can be used to express the $z_i$'s as a $\sqrt{T}$-expansion whose coefficients are functions on the spectral curve that also depend on the coefficients of the potentials. 
\beq
z_i(p)=\sum_{n=1}^\infty z_i^{(n/2)}(g_3^{(1)},g_3^{(2)},g_3^{(3)};p)T^{n/2}
\eeq
For example, the first few orders of $z_1(p)$ are given by
\beq
z_1^{(1/2)}\left(g_3^{(j)};p\right)=-\left(p+\frac{2}{p}\right),
\eeq
which can be recognized as the gaussian contribution,
\beq
z_1^{(1)}\left(g_3^{(j)}; p\right)=\frac{g_3^{(2)}+3g_3^{(3)}}{p^2}-8g_3^{(1)}-2g_3^{(2)}-4g_3^{(3)},
\eeq
\begin{eqnarray*}
z_1^{(3/2)}\left(g_3^{(j)}; p\right)&=& p \left(-16(g_3^{(1)})^2-7g_3^{(1)}g_3^{(2)}-2(g_3^{(2)})^2-13g_3^{(1)}g_3^{(3)}-7g_3^{(2)}g_3^{(3)}-16(g_3^{(3)})^2\right) \\
          &+&\frac{2g_3^{(2)}g_3^{(3)}}{p^3}+\frac{-32g_3^{(1)}-6g_3^{(1)}g_3^{(2)}-2g_3^{(1)}g_3^{(3)}+16(g_3^{(3)})^2}{p},
\end{eqnarray*}

\begin{eqnarray*}
z_1^{(2)}\left(g_3^{(j)}; p\right)&=& -384(g_3^{(1)})^3-192g_3^{(2)}(g_3^{(1)})^2-60g_3^{(1)}(g_3^{(2)})^2-12(g_3^{(2)})^3-288(g_3^{(1)})^2g_3^{(3)}-
\\&-& 162g_3^{(1)}g_3^{(2)}g_3^{(3)}-48(g_3^{(2)})^2g_3^{(3)}-216g_3^{(1)}(g_3^{(3)})^2-120g_3^{(2)}(g_3^{(3)})^2+192(g_3^{(3)})^3\\
&+&\frac{1}{p^4}g_3^{(2)}(g_3^{(3)})^2+\frac{1}{p^2}\left(32(g_3^{(1)})^2g_3^{(2)} 14g_3^{(1)}(g_3^{(2)})^2+4(g_3^{(2)})^3+96(g_3^{(1)})^2g_3^{(3)}\right),
\end{eqnarray*}
 and so on.

 The same treatment can be given to the other meromorphic parametrizations and obtain similar expressions :
\beq
z_2(p)=-\sqrt{T}\left(p+\frac{1}{p}\right)+T\left(\frac{g_3^{(3)}} {p^2}+g_3^{(1)}p^2-4g_3^{(1)}-2g_3^{(2)}-4g_3^{(3)}\right)
+...
\eeq
\beq
z_3(p)=-\sqrt{T}\left(2p+\frac{1}{p} \right)+T\left(\left(3g_3^{(1)}+g_3^{(2)}\right)p^2-8g_3^{(1)}-2g_3^{(2)}-4g_3^{(3)}\right)+...
\eeq

\begin{figure}
  \begin{center}
  \includegraphics[width=12cm,angle=00]{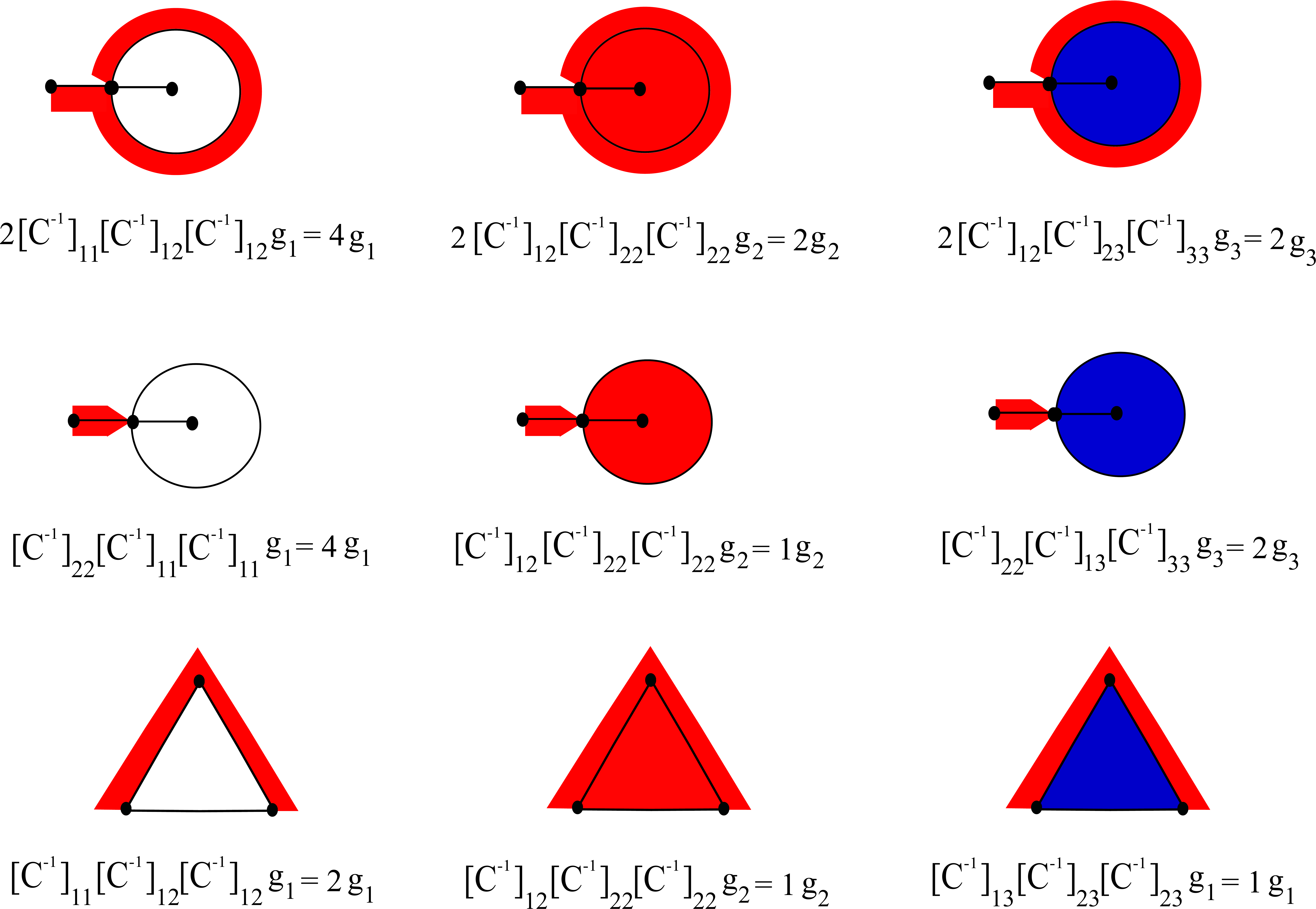}\\
  \end{center}
  \caption{ Maps contributing to ${\cal T}_{1,2,0}^2$. The outer face is the marked face whose boundary has type (1,2,2).}\label{3vertex}
\end{figure}

It is important to notice that these coefficients have to be found only once and thereafter they can be used to compute the $T$- expansions for the correlation functions without any further modification. Using these parametrizations and  \eq{3mm}, we compute order by order the terms in the formal  $T$-expansion for the disk amplitude with mixed boundary conditions i.e.
\beq
W_{1,2,3}(z_1,z_2,z_3)=\sum_{v=0}^\infty\sum_{\vec{n}\in\mathbb{N}^3}\frac{\mathcal{T}_{\vec{n}}^v(\vec{g_3})}{z_1^{n_1+1}z_2^{n_2+1}z_3^{n_3+1}}T^v.
\eeq
It is important to recall that
\beq
\sum_{v=0}^\infty\mathcal{T}_{\vec{n}}^v(\vec{g_3})T^v=\lim_{N\to\infty} {1 \over N} \moy{Tr\left(M_1^{n_1}M_2^{n_2}M_3^{n_3}\right)}.
\eeq

Here $\mathcal{T}^v_{\vec{n}}(\vec{g_3})$ denotes the contributions to the correlation function due to disks with $v$ vertices and whose boundary conditions are given by an ordered sequence $(n_1,n_2, n_3)$ of boundaries of types $1,2$ and $3$ respectively. Since $g_2^{(1)}=g_2^{(3)}$ the symmetry
\beq
\mathcal{T}^v_{(n_1,n_2,n_3)}(g^{(1)}_3,g^{(2)}_3,g^{(3)}_3)=\mathcal{T}^v_{(n_3,n_2,n_1)}(g^{(3)}_3,g^{(2)}_3,g^{(1)}_3)
\eeq
holds. Moreover,  the $\mathcal{T}^v_{\vec{n}}(\vec{g_3})$'s are given by power expansions in the couplings of the cubic terms in the potentials, i.e
\beq
\mathcal{T}^v_{\vec{n}}(\vec{g_3})=\sum_{\vec{m}\in\mathbb{N}^3}\mathcal{T}^v_{{\vec{n}},{\vec{m}}}(g_3^{(1)})^{m_1}(g_3^{(2)})^{m_2}(g_3^{(3)})^{m_3},
\eeq
here $\mathcal{T}^v_{{\vec{n}},{\vec{m}}}$ has to be understood as the sum of the contributions to $\mathcal{T}^v_{{\vec{n}}}$ containing $m_i$ triangles of type $i=1,2,3$.  It is worth seeing this explicitly for a couple of simple examples:

\beq
\mathcal{T}_{(1,0,0)}^2(\vec{g_3})=-\left(4g_3^{(1)}+g_3^{(2)}+2g_3^{(3)}\right)
\eeq

\beq
\mathcal{T}_{(1,2,0)}^3(\vec{g_3})=-\left(10g_3^{(1)}+4g_3^{(2)}+7g_3^{(3)}\right)
\eeq
The explicit enumeration of maps giving rise to the coefficients of these numbers is presented in fig.\ref{2vertex} and fig.\ref{3vertex}.

The contributions to $W_{1,2,3}$ up to three vertices  (omitting terms that can be obtained by symmetry) are displayed in the following table. For clarity we 
use the simplified notation $g_3^{(1)}=g_1$, $g_3^{(2)}=g_2$ and $g_3^{(3)}=g_3$.

\begin{center}
\begin{tabular}{ | l || p{2cm} | p{3cm}| p{4.5cm} | }
  \hline
  Boundaries $\backslash$ Vertices & 1 & 2 &3\\ \hline
  $(1,0,0)$&1&$-4g_3-g_2-2g_3$&$-128g_1^3$ $-64g_1^2g_2$ $-20g_1g_2^2$ $-4g_2^3$ $-96g_1^2g_3$ $-54g_1g_2g_3$ $-16g_2^2g_3$ $-72g_1g_3^2$ $-40g_2g_3^2$ $-64g_3^3$\\  \hline
  $(2,0,0)$&0&2&$64g_1^2$ $+24g_1g_2$ $+4g_2^2$ $+40g_1g_3$ $+12g_2g_3$  $+16g_3^2$\\  \hline
  $(3,0,0)$&0&0&$-32g1$ $-7g_2$ $-13g_3$\\  \hline
  $(4,0,0)$&0&0&$8$\\  \hline
  
   $(0,1,0)$&1&$-2g_1-g_2-2g_3$& $-64g_1^3$ $-40g_1^2g_3^2$ $-16g_1g_2^2$ $-4g_2^3$ $-56g_1^2g_3$ $-42g_1g_2g_3$ $-16g_2^2g_3$ $-56g_1g_3^2$ $-40g_2g_3^2$ $-64g_3^2$\\  \hline
   $(0,2,0)$&0&1&$16g_1^3 $ $+12g_1g_2$ $+4g_2^2$ $+18g_1g_3$  $+12g_2g_3$ $+16g_3^3$\\  \hline
   $(0,3,0)$&0&0&$-7g_1$ $-4g_2$ $-7g_3$\\  \hline
   $(0,4,0)$&0&0&2\\  \hline

    $(1,1,0)$&0&1&$32g_1^2$ $+17g_1g_2$ $+4g_2^2$ $+27g_1g_3$ $+12g_2g_3$ $+16g_3^2$\\  \hline
    $(1,2,0)$&0&0&$-10g_1$ $-4g_2$ $-7g_3$\\  \hline
    $(1,3,0)$&0&0&2\\  \hline
    $(2,1,0)$&0&0&$-16g_1$ $-5g_2$ $-9g_3$\\  \hline
    $(3,1,0)$&0&0&4\\  \hline
    $(2,2,0)$&0&0&1\\  \hline

    $(1,0,1)$&0&0&$-32g_1^2$ $-19g_1g_2$ $-5g_2^2$ $-41g_1g_3$ $-19g_2g_3$ $-32g_3^2$\\  \hline
    $(2,0,1)$&0&-1&$16g_1$ $+7g_2$ $+14g_3$\\  \hline
    $(3,0,1)$&0&0&-4 \\  \hline
    $(2,0,2)$&0&0&7\\  \hline
     
     $(1,1,1)$&0&0&$-2g_1$ $-g_2$ $-4g_3$\\  \hline
     $(2,1,1)$&0&0&1\\  \hline
     $(1,2,1)$&0&0&2\\  \hline

   \end{tabular}
\end{center}

\bigskip

\section{Conclusion and prospects}

In these notes, we found a recursive formula giving access to the so called complete disk amplitude in the multi-matrix model where an arbitrary number of matrices are coupled in chain. As in the two matrix model, this recursive formula only involves the spectral curve of the matrix model. Once the latter is known by the computation of the non-mixed disk amplitude, the recursive procedure presented in this paper explains how to get a disk with mixed boundary conditions by simple series expansion in terms of local variables on the spectral curve.
This result proves a conjecture of Eynard \cite{Echain} and is the first step towards the generalization of the topological recursion formalism for the computation of mixed amplitude in the mutli-matrix setup. From the two matrix model experience \cite{EOallmixed}, it was also anticipated that the mixed correlation functions should follow from recursion relations taking into account all possible degenerations of the maps enumerated.

This work obviously calls for generalizations and the computation of arbitrary mixed amplitude to any order in the $1/N^2$ expansion will probably be accessible by a recursive procedure similar to this one. This problem will be addressed in a forthcoming paper.

On the other hand,  a direct combinatorial understanding of these recursion relations is still missing and would deserve further investigations. Can we give any meaning to the so-called non-physical sheets of the spectral curve? Can we get a universal factorization formula for arbitrary disk amplitude as in \cite{EObethe}? These problems are definitely of high interest, not only for combinatorial motivations but also to investigate the generic structure underlying the mixed amplitudes in matrix models. This last point might be fundamental to understand the possible applications of these amplitudes to fields other than random matrix theories, such as topological string theories.

\bigskip

\bigskip

\bigskip

\noindent {\Large \bf Aknowledgement}

N. O. would like to thank Bertrand Eynard and Aleix Prats-Ferrer for fruitful discussions on the subject.
The work of the author  is founded by  the Funda\c{c}\~ao para a Ci\^{e}ncia e a Tecnologia through the fellowship SFRH/BPD/70371/2010 for N.O and SFRH / BD / 64446 / 2009
for A. V.-O.

%%%%%%%%%%%%%%%%%%%%%%%%%%%%%%%%%%%%%%%%%%%%%%%%
%%%%%%%%%%%%%%%%%%%%%%%%%%%%%%%%%%%%%%%%%%%%%%%%
%%%%%%APPENDIX
%%%%%%%%%%%%%%%%%%%%%%%%%%%%%%%%%%%%%%%%%%%%%%%%%%%%%%%%%%%%%%%%%%%%%%%%%%%%%%%%%%%%%%%%%%%%%%%%
\newpage

\setcounter{section}{0}
\appendix{Derivation of the loop equations}

\subsection{Loop equation for the complete mixed function}

Let us recall the derivation of the loop equations in the setup of \cite{Echain}.

Consider the change of variable $\delta M_1 = {1 \over x_2 - M_2} \dots {1 \over x_\CN - M_\CN} {1 \over x_1-M_1}$. The associated loop equation reads
\bea
&& \left[c_{1,2}x_2 - Y(x_1)\right] W_{1,\dots,\CN}(x_1,\dots,x_\CN) - {T \over N^2} W_{1,\dots \CN;1}(x_1,\dots,x_\CN;x_1) = \cr
&& = - P_{1,1}(x_1,\dots,x_\CN) + c_{1,2}W_{1,3,4,\dots,\CN}(x_1,x_3,\dots,x_\CN). \cr
\eea

One can extract some interesting information from this equation. It can indeed read
\bea
\Res_{x_2 \to \infty} W_{1,\dots,\CN}(x_1,\dots,x_\CN) V_2'(x_2) dx_2 &=&
- \Res_{x_2 \to \infty} {P_{1,1}(x_1,\dots,x_\CN)  V_2'(x_2) dx_2 \over c_{1,2}x_2 - Y(x_1)} \cr
&& + c_{1,2}  \Res_{x_2 \to \infty} {W_{1,3,4,\dots,\CN}(x_1,x_3,\dots,x_\CN)  V_2'(x_2) dx_2 \over c_{1,2}x_2 - Y(x_1)} \cr
&& + {T \over N^2} \Res_{x_2 \to \infty} {W_{1,\dots \CN;1}(x_1,\dots,x_\CN;x_1)  V_2'(x_2) dx_2 \over c_{1,2}x_2 - Y(x_1)} .\cr
\eea 
One can remark that, for a function $f(x)$ analytic in $x \to \infty$
\beq
\Res_{x \to \infty} {f(x) dx \over x-x_1} = \Pol_{x_1} f(x_1)
\eeq
which implies that
\bea
c_{1,2} \Res_{x_2 \to \infty} W_{1,\dots,\CN}(x_1,\dots,x_\CN) V_2'(x_2) dx_2 &=& c_{1,2} V_2'(\hat{x}_2) W_{1,3,4,\dots,\CN}(x_1,x_3,\dots,x_\CN) \cr
&&+ \left[ {T \over N^2}  \Pol_{x_2} W_{1,\dots \CN;1}(x_1,\dots,x_\CN;x_1)  V_2'(x_2) \right.  \cr
&& \quad  - \left. \Pol_{x_2} P_{1,1}(x_1,\dots,x_\CN) V_2'(x_2)\right]_{c_{1,2} x_2 = Y(x_1)} . \cr
\eea
Using the notations of the preceding section, one gets
\bea\label{eqinterres}
c_{1,2} \Res_{x_2 \to \infty} W_{1,\dots,\CN}(x_1,\dots,x_\CN) V_2'(x_2) dx_2 &=&
- P_{1,2}(x_1,\hat{x}_2,x_3,\dots,x_\CN) \cr
&& - c_{1,2} W_{3,4,\dots,\CN}(x_3,\dots,x_\CN) \cr
&&+  c_{1,2} V_2'(\hat{x}_2) W_{1,3,4,\dots,\CN}(x_1,x_3,\dots,x_\CN) \cr
&&+ {T \over N^2}  P_{2,2;1}(x_1,\hat{x}_2,x_3,\dots,x_\CN) .\cr
\eea

%\br
%Forgetting about the notations, it reads
%\bea
%&& \left[c_{1,2}x_2 - Y(x_1)\right] W_{1,\dots,\CN}(x_1,\dots,x_\CN) - {T \over N^2} W_{1,\dots \CN;1}(x_1,\dots,x_\CN;x_1) = \cr
%&& = - \Pol_{x_1} V_1'(x_1) W_{1,\dots,\CN}(x_1,\dots,x_\CN) + c_{1,2}W_{1,3,4,\dots,\CN}(x_1,x_3,\dots,x_\CN) \cr
%\eea
%and
%\bea
%c_{1,2} \Res_{x_2 \to \infty} W_{1,\dots,\CN}(x_1,\dots,x_/CN) V_2'(x_2) dx_2 &=&
%- \left[\Pol_{x_1,x_2} V_1'(x_1) V_2'(x_2) W_{1,\dots,\CN}(x_1,\dots,x_\CN) \right]_{c_{1,2} x_2 = Y(x_1)} \cr
%&&+  c_{1,2} V_2'(\hat{x}_2) W_{1,3,4,\dots,\CN}(x_1,x_3,\dots,x_\CN) \cr
%&&+ {T \over N^2}  \left[\Pol_{x_2} W_{1,\dots \CN;1}(x_1,\dots,x_\CN;x_1)  V_2'(x_2)\right]_{c_{1,2} x_2 = Y(x_1)} .\cr
%\eea
%
%
%
%\er

\subsection{Hierarchy of loop equations}

Let us now consider the change of variable $\delta M_2 = {1 \over x_3-M_3} {1 \over x_4-M_4} \dots {1 \over x_1 - M_1}$. The corresponding loop equation is
\bea
0 &=& {T \over N} \left< \Tr {1 \over x_1-M_1} V_2'(M_2)  {1 \over x_3-M_3} \dots {1 \over x_\CN-M_\CN}\right> + c_{1,2} W_{3,4,\dots,\CN}(x_3,\dots,x_\CN)  \cr
&& + c_{2,3} W_{1,4,\dots,\CN} - \left(c_{1,2} x_1 + c_{2,3} x_3\right) W_{1,3,4,\dots,\CN}(x_1,x_3,\dots,x_\CN) . \cr
\eea
The first term is actually
\beq
{T \over N} \left< \Tr {1 \over x_1-M_1} V_2'(M_2)  {1 \over x_3-M_3} \dots {1 \over x_\CN-M_\CN}\right>   =  \Res_{x_2 \to \infty} W_{1,\dots,\CN}(x_1,\dots,x_\CN) V_2'(x_2) dx_2
\eeq
which implies, using \eq{eqinterres},
%\br
%Without notations, one gets
%\bea
%0 &=&- \left[\Pol_{x_1,x_2} {V_1'(x_1) V_2'(x_2) \over c_{1,2}} W_{1,\dots,\CN}(x_1,\dots,x_\CN) \right]_{c_{1,2} x_2 = Y(x_1)} \cr
%&&+  V_2'(\hat{x}_2) W_{1,3,4,\dots,\CN}(x_1,x_3,\dots,x_\CN) \cr
%&&+ {T \over N^2}  \left[\Pol_{x_2} W_{1,\dots \CN;1}(x_1,\dots,x_\CN;x_1)  {V_2'(x_2) \over c_{1,2}}\right]_{c_{1,2} x_2 = Y(x_1)} + c_{1,2} W_{3,4,\dots,\CN}(x_3,\dots,x_\CN) + c_{2,3} W_{1,4,\dots,\CN}(x_1,x_4,\dots,x_\CN) \cr
%&& - \left(c_{1,2} x_1 + c_{2,3} x_3\right) W_{1,3,4,\dots,\CN}(x_1,x_3,\dots,x_\CN) . \cr
%\eea
%This gives
\bea
\left(x_3 - \hat{x}_3(x_1)\right) W_{1,3,\dots,\CN}(x_1,x_3,\dots,x_\CN) &=& W_{1,4,\dots,\CN}(x_1,x_4,\dots,x_\CN) \cr
&& - \left[\Pol_{x_1,x_2} {V_1'(x_1) V_2'(X_2) \over c_{1,2} c_{1,3}}  W_{1,\dots,\CN}(x_1,\dots,x_\CN)\right. \cr
&& \quad + \Pol_{x_1,x_2}  {c_{1,2} x_1 x_2 \over c_{2,3}} W_{1,\dots,\CN}(x_1,\dots,x_\CN) \cr
&& \quad + \left. {T \over N^2} \Pol_{x_2} {V_2'(x_2) \over c_{1,2} c_{2,3}} W_{1,\dots,\CN}(x_1,\dots,x_\CN)\right]_{x_2 = \hat{x}_2}\cr
\eea
where
\beq
\hat{x}_3 := {V_2'(\hat{x}_2) - c_{1,2} x_1 \over c_{2,3}}.
\eeq

Let us now proceed by induction.  Let $0<k<\CN$
and assume that one has
\bea
&& \left(x_j - \hat{x}_j(x_1)\right) W_{1,j,\dots,\CN}(x_1,x_j,\dots,x_\CN) = \cr
&&  \qquad \qquad \qquad \qquad =  W_{1,j+1,\dots,\CN}(x_1,x_{j+1},\dots,x_\CN) \cr
&&  \qquad \qquad \qquad \qquad - \left[\Pol_{x_1,\dots,x_{j-1}}f_{1,j-1}(x_1,\dots,x_{j-1}) W_{1,\dots,\CN}(x_1,\dots,x_\CN)\right]_{\left\{x_i = \hat{x}_i\right\}_{i = 2}^{j-1}} \cr
&& \qquad \qquad   \qquad \qquad + {T \over N^2} \left[\Pol_{x_2,\dots,x_{j-1}} f_{2,j-1}(x_2,\dots,x_{j-1}) W_{1,\dots,\CN}(x_1,\dots,x_\CN)\right]_{\left\{x_i = \hat{x}_i\right\}_{i = 2}^{j-1}} \cr
\eea
for $j\leq k$. We now show that it is also true for $j=k+1$.

This property for $j=k$ implies that
\bea
 && \Res_{x_k} V_k'(x_k) dx_k W_{1,k,\dots,\CN}(x_1,x_k,\dots,x_\CN)  = \cr
&&   \Res_{x_k} {V_k'(x_k) dx_k  W_{1,k+1,\dots,\CN}(x_1,x_{k+1},\dots,x_\CN) \over \left(x_k - \hat{x}_k(x_1)\right)}\cr
&& - \Res_{x_k} {V_k'(x_k) dx_k \left[\Pol_{x_1,\dots,x_{k-1}}f_{1,k-1}(x_1,\dots,x_{k-1}) W_{1,\dots,\CN}(x_1,\dots,x_\CN)\right]_{\left\{x_i = \hat{x}_i\right\}_{i = 2}^{k-1}} \over \left(x_k - \hat{x}_k(x_1)\right)}\cr
&& + {T \over N^2} \Res_{x_k} {V_k'(x_k) dx_k  \left[\Pol_{x_2,\dots,x_{k-1}} f_{2,k-1}(x_2,\dots,x_{k-1}) W_{1,\dots,\CN}(x_1,\dots,x_\CN)\right]_{\left\{x_i = \hat{x}_i\right\}_{i = 2}^{k-1}} \over \left(x_k - \hat{x}_k(x_1)\right)}\cr
\eea
i.e.
\bea
 && \Res_{x_k} V_k'(x_k) dx_k W_{1,k,\dots,\CN}(x_1,x_k,\dots,x_\CN)  = \cr
&& V_k'(\hat{x}_k) W_{1,k+1,\dots,\CN}(x_1,x_{k+1},\dots,x_\CN) \cr
&& - \left[\Pol_{x_1,\dots,x_{k}}  V_k'(x_k)  f_{1,k-1}(x_1,\dots,x_{k-1}) W_{1,\dots,\CN}(x_1,\dots,x_\CN)\right]_{\left\{x_i = \hat{x}_i\right\}_{i = 2}^{k}} \cr
&& + {T \over N^2}  \left[\Pol_{x_2,\dots,x_{k}} V_k'(x_k)f_{2,k-1}(x_2,\dots,x_{k-1}) W_{1,\dots,\CN}(x_1,\dots,x_\CN)\right]_{\left\{x_i = \hat{x}_i\right\}_{i = 2}^{k}} .\cr
\eea
 
 Let us consider the induction hypothesis for $j=k-1$, it gives
 \bea
 && \Res_{x_{k-1}} dx_{k-1} x_{k-1} W_{1,k-1,\dots,\CN }(x_1,x_{k-1},\dots,x_\CN) = \cr
 && \hat{x}_{k-1} W_{1,k,\dots,\CN }(x_1,x_{k},\dots,x_\CN) \cr
 && - \left[\Pol_{x_1,\dots,x_{k-1}} x_{k-1} f_{1,k-2}(x_1,\dots,x_{k-1}) W_{1,\dots,\CN}(x_1,\dots,x_\CN)\right]_{\left\{x_i = \hat{x}_i\right\}_{i = 2}^{k-1}} \cr
 && + {T \over N^2} \left[\Pol_{x_2,\dots,x_{k-1}} x_{k-1} f_{2,k-2}(x_1,\dots,x_{k-1}) W_{1,\dots,\CN}(x_1,\dots,x_\CN)\right]_{\left\{x_i = \hat{x}_i\right\}_{i = 2}^{k-1}} .\cr
 \eea

The loop equation corresponding to $\delta  M_k = {1 \over x_{k+1}- M_{k+1}} \dots {1 \over x_\CN - M_\CN} {1 \over x_1 - M_1}$ reads (for $k<\CN$)
\bea
0 &=& {T \over N} \left< \Tr {1 \over x_1-M_1} V_k'(M_k)  {1 \over x_{k+1}-M_{k+1}} \dots {1 \over x_\CN-M_\CN}\right> \cr
&& - c_{k-1,k} \Res_{x_{k-1},x_k}dx_k dx_{k-1} x_{k-1} W_{1,k-1,\dots,\CN}(x_1,x_{k-1},\dots,x_\CN)\cr
&& 
+ c_{k,k+1} W_{1,k+2,\dots,\CN}(x_1,x_{k+2},\dots,x_\CN) \cr
&& - c_{k,k+1} x_{k+1} W_{1,k+1,\dots,\CN}(x_1,x_{k+1},\dots,x_\CN) . \cr
\eea

 Plugging in the expression found earlier, one gets
\bea
&& \left(x_{k+1} - \hat{x}_{k+1}(x_1)\right) W_{1,k+1,\dots,\CN}(x_1,x_{k+1},\dots,x_\CN) = \cr
&&  \qquad \qquad  = W_{1,k+2,\dots,\CN}(x_1,x_{k+1},\dots,x_\CN) \cr
&&  \qquad \qquad - \left[\Pol_{x_1,\dots,x_{k-1}}f_{1,k}(x_1,\dots,x_{k-1}) W_{1,\dots,\CN}(x_1,\dots,x_\CN)\right]_{\left\{x_i = \hat{x}_i\right\}_{i = 2}^{k-1}} \cr
&&  \qquad \qquad + {T \over N^2} \left[\Pol_{x_2,\dots,x_{k-1}} f_{2,k}(x_2,\dots,x_{k-1}) W_{1,\dots,\CN}(x_1,\dots,x_\CN)\right]_{\left\{x_i = \hat{x}_i\right\}_{i = 2}^{k-1}} \cr
\eea 
 with
 \beq
 \hat{x}_{k+1} = {V_k'(\hat{x}_k) - c_{k-1,k} \hat{x}_{k-1} \over c_{k,k+1}},
 \eeq
 \beq
 f_{1,k}(x_1,\dots,x_{k-1}) = {V_k'(x_k) f_{1,k-1} - c_{k-1,k} x_{k-1} x_k f_{1,k-2}\over c_{k,k+1}}
 \eeq
 and
 \beq
 f_{2,k}(x_1,\dots,x_{k-1}) = {V_k'(x_k) f_{2,k-1} - c_{k-1,k} x_{k-1} x_k f_{2,k-2}\over c_{k,k+1}}.
 \eeq
 Together with the initial equation for $W_{1,\CN}(x_1,x_\CN)$, this ends the proof.


\begin{thebibliography}{110}


%%%%%%%%%%%%%%%%%%%%%%%%%%%%%%%%%%%%%%%%%%%%%%%%%%%%%%%%%%%%%%
%%%%%%%%Modeles de matrices


\bibitem{ACKM}
J.Ambj{\o}rn, L.Chekhov, C.F.Kristjansen and Yu.Makeenko,
``Matrix model calculations beyond the spherical limit'',
{\em Nucl.Phys.} {\bf B404} (1993) 127--172; Erratum ibid. {\bf B449} (1995) 681,
hep-th/9302014.

\bibitem{ambjornrmt} J. Ambj{\o}rn, B. Durhuus, J. Fr{\"o}hlich, "Diseases of triangulated random surface models, and possible cures", {\em Nucl. Phys.} {\bf B257}, p. 433-449.


%\bibitem{Banks} T.Banks, M.R.Douglas, N.Seiberg, S.H.Shenker, {\em Phys. Lett.} {\bf B238}(1990)279.

\bibitem{BEMPF} M. Berg\`ere, B. Eynard, O. Marchal, A. Prats-Ferrer, ``Loop equations and topological recursion for the arbitrary-$\beta$ two-matrix model'',
 arXiv:1106.0332.




\bibitem{BIPZ} E. Br\'ezin, C. Itzykson, G. Parisi, and J.B. Zuber,
{\em Comm. Math. Phys.} {\bf 59}, 35  (1978).




\bibitem{ec1loopF} L.Chekhov, B.Eynard,
``Hermitian matrix model free energy: Feynman graph technique for all genera'',
{\em J. High Energy Phys.} {\bf JHEP03} (2006) 014, hep-th/0504116.

\bibitem{ChEynbeta} L.Chekhov, B.Eynard,
``Matrix eigenvalue model: Feynman graph technique for all genera'',
{\em J. High Energy Phys.} {\bf JHEP 0612} (2006) 026, math-ph/0604014.


\bibitem{CEM1} L. O. Chekhov, B. Eynard, O. Marchal, ``Topological expansion of beta-ensemble model and quantum algebraic geometry in the sectorwise approach'',
{\em Theor.Math.Phys.} 166 (2011) 141, arXiv:1009.6007.


\bibitem{CEM2} L. Chekhov, B. Eynard, O. Marchal, ``Topological expansion of the Bethe ansatz, and quantum algebraic geometry'',
arXiv:0911.1664.




\bibitem{CEO} L.Chekhov, B.Eynard and N.Orantin,
``Free energy topological expansion for the 2-matrix model'',
{\em J. High Energy Phys.} {\bf JHEP12} (2006) 053, math-ph/0603003.




\bibitem{David} F. David, "Planar diagrams, two-dimensional lattice gravity and surface models",
{\em Nucl.Phys.} {\bf B257}(1985)45.







\bibitem{E1MM} B. Eynard, ``Topological expansion for the 1-hermitian matrix model correlation functions'',
JHEP/024A/0904, hep-th/0407261.

\bibitem{eynm2m} B. Eynard, ``Large N expansion of the 2-matrix model'',
{\em J. High Energy Phys.} {\bf JHEP01} (2003) 051, hep-th/0210047.

\bibitem{Echain} B. Eynard, ``Master loop equations, free energy and correlations for the chain of matrices'',
{\em J. High Energy Phys.}  {\bf JHEP11}(2003)018, hep-th/0309036, ccsd-00000572.






\bibitem{Emap} B.Eynard,
`` Formal matrix integrals and combinatorics of maps'', \\
math-ph/0611087.

\bibitem{EKK}
B.~Eynard, A.~Kokotov, and D.~Korotkin,
``$1/N^2$ corrections to free energy in Hermitian two-matrix model'',
hep-th/0401166.

\bibitem{EM} B. Eynard, O. Marchal, ``Topological expansion of the Bethe ansatz, and non-commutative algebraic geometry'', {\em J. High Energy Phys.} {\bf JHEP 0903}(2009)094,
arXiv:0809.3367.

\bibitem{EObethe} B.Eynard, N.Orantin,``Mixed correlation functions in the 2-matrix model, and
the Bethe Ansatz�, {\em J. High Energy Phys.} {\bf JHEP08}(2005)028, math-ph/0504029.

\bibitem{EO1} B.Eynard, N.Orantin,
``Topological expansion of the 2-matrix model correlation functions: diagrammatic rules for a residue formula'',
{\em J. High Energy Phys.} {\bf JHEP12}(2005)034, math-ph/0504058.

\bibitem{EOallmixed} B. Eynard, N. Orantin, ``Topological expansion and boundary conditions''
{\em J. High Energy Phys.} {\bf JHEP06} (2008) 037, arXiv:0710.0223.



\bibitem{EOinvariants} B.Eynard, N.Orantin,
``Invariants of algebraic curves and topological expansion'', {\em Communications in Number Theory and Physics,} Vol 1, Number 2, math-ph/0702045.

\bibitem{EOsymmetry} B.Eynard, N.Orantin, ``Topological expansion of mixed correlations in the hermitian 2 Matrix Model and x-y symmetry of the Fg invariants'',
{\em J. Phys. Math. Theor.}  {\bf A41} (2008) 015203,arXiv:0705.0958.

\bibitem{EOrev} B.Eynard, N.Orantin, ``Topological recursion in enumerative geometry and random matrices'',
{\em J. Phys. A: Math. Theor.} {\bf 42} 293001,
arXiv:0811.3531.



\bibitem{EPF} B. Eynard, A. Prats-Ferrer, ``Topological expansion of the chain of matrices'',
{\em J. High Energy Phys.} {\bf JHEP07}(2009)096,
arXiv:0805.1368.



\bibitem{Migdal} D. Gross, A. Migdal, {\em Phys. Rev. Lett.} {\bf 64}(1990)127; {\em Nucl. Phys.} {\bf B340}(1990)333.



\bibitem{Kazakov} V.A. Kazakov, "Bilocal regularization of models of random surfaces"
{\em Physics Letters B} {\bf 150}(1985), Issue 4, p. 282-284.




\bibitem{thooft} G. 't Hooft, {\em Nuc. Phys.} {\bf B72}, 461 (1974).


\bibitem{tutte} W.T. Tutte, ``A census of planar triangulations'',
{\em Can. J. Math.} {\bf 14} (1962) 21-38.

\bibitem{tutte2} W.T. Tutte, ``A census of planar maps'',
{\em Can. J. Math.} {\bf 15} (1963) 249-271.


















\end{thebibliography}
\end{document}